\documentclass[pra,preprint,11pt,superscriptaddress]{revtex4-1} 
\usepackage{amsmath}
\usepackage{amssymb}
\usepackage{graphicx}
\usepackage{color}
\newcommand{\be}[1]{\begin{equation}\label{#1}}
\newcommand{\ee}{\end{equation}}
\newcommand{\bea}[1]{\begin{eqnarray}\label{#1}}
\newcommand{\eea}{\end{eqnarray}}
\newcommand{\no}{\nonumber \\}
\newcommand{\Fig}[1]{Fig.(\ref{#1})}
\newcommand{\Tbl}[1]{Table \ref{#1}}
\newcommand{\Eq}[1]{Eq.(\ref{#1})}
\newcommand{\App}[1]{Appendix~\ref{#1}}
\newcommand{\Sec}[1]{Section~\ref{#1}}
\newcommand{\bsub}{\begin{subequations}}
\newcommand{\esub}{\end{subequations}}

\usepackage{color}

\def\trm#1{\textrm{#1}}
\def\tit#1{\textit{#1}}

\def\a0{{\alpha_0}}

\def\da0{{\dot{\alpha}_0}}

\def\myover#1{\myoverDefn#1}
\def\myoverDefn#1#2{\hbox{\space \raise-2mm\hbox{$\textstyle{#1} \atop \scriptstyle{#2}$} }}

\def\t{{\tau}}

\def\a{{\alpha}}

\def\dag{\dagger}

\def\S{\mathcal{S}}
\def\B{\mathcal{B}}
\def\U{\mathcal{U}}

\def\Ucc{\mathcal{U}^{(cc)}}
\def\Ucs{\mathcal{U}^{(cs)}}
\def\Uss{\mathcal{U}^{(ss)}}

\def\Uccprime{\mathcal{U}^{\prime (cc)}}
\def\Ucsprime{\mathcal{U}^{\prime (cs)}}
\def\Ussprime{\mathcal{U}^{\prime (ss)}}
\def\Ucprime{\mathcal{U}^{\prime (c)}}
\def\Usprime{\mathcal{U}^{\prime (s)}}

\def\rp{r_{P}}
\def\rp2{r_{p}^{2}}

\newcommand{\ket}[1]{|#1\rangle}
\newcommand{\bra}[1]{\langle #1|}

\newcommand{\Order}[1]{\mathcal{O}(#1)}
\newcommand{\IP}[2]{\langle {#1} | {#2} \rangle}

\begin{document}
\title{A direct interferometric test of the nonlinear phase shift gate}
\author{Peter L. Kaulfuss}
\affiliation{Rochester Institute of Technology, School of Physics and Astronomy, 85 Lomb Memorial Dr., Rochester, NY 14623}
\author{Paul M. Alsing}\email{corresponding author: paul.alsing@us.af.mil}
\affiliation{Air Force Research Laboratory, Information Directorate, 525 Brooks Rd, Rome, NY, 13411}
\author{Edwin~E.~Hach~III}
\affiliation{Rochester Institute of Technology, School of Physics and Astronomy, 85 Lomb Memorial Dr., Rochester, NY 14623}
\author{A.~Matthew~Smith}\author{Michael L. Fanto}
\affiliation{Air Force Research Laboratory, Information Directorate, 525 Brooks Rd, Rome, NY, 13411}

\date{\today}

\begin{abstract}
We propose a direct interferometric test of the Non-Linear Phase Shift Gate (NLPSG), an essential piece of a Knill Laflamme Milburn Contolled-NOT (KLM CNOT) gate. We develop our analysis for the both the case of the original, bulk optical KLM NLPSG and for the scalable integrated nano-photonic NLPSG based on Micro-Ring Resonators (MRRs) that we have proposed very recently. Specifically, we consider the interference between the target photon mode of the NLPSG along one arm of a Mach Zehnder Interferometer (MZI) and a mode subject to an adjustable linear phase along the other arm. Analysis of triple-photon coincidences between the two modes at the output of the MZI and the success ancillary mode of the NLPSG provides a signature of the 
operation of the NLPSG. We examine the triple coincidence results for experimentally realistic cases of click/no-click detection with sub-unity detection efficiencies. Further we compare the case for which the MZI input modes are seeded with weak Coherent States (w-CS) and to that for which the input states are those resulting from colinear Spontaneous Parametric Down Conversion (cl-SPDC). In particular, we show that, though more difficult to prepare, cl-SPDC states offer clear advantages for performing the test, especially in the case of relatively low photon detector efficiency.
\end{abstract}
\maketitle

\section{Introduction}\label{sec:Intro}
In 2001, Knill, Laflamme, and Milburn (KLM) presented an efficient for linear optical quantum computing based on a probabilistic Controlled-NOT (CNOT) gate \cite{KLM:2001} using 3 photons. 
The following year, Knill \cite{Knill:2002} improved the KLM-CNOT success probability from $1/16$ to $2/27$ 
by means of a 4-photon conditional sign shift gate.
A decade later, Okamoto et. al. demonstrated an experimental realization of the KLM CNOT \cite{Obrien:2011}. The original realization of the KLM CNOT was in a bulk optical setting, and, therefore, was not scalable as would be required for the deployment of the gate as part of a practical computing system. So, nearly a decade after that, the present authors proposed a scalable version of the KLM CNOT based upon an integrated silicon nano-photonic architecture composed of directionally coupled Micro-Ring Resonators (MRRs) \cite{Alsing_Hach:2019, Alsing_Hach:2018}. Currently, we are in the early stages of an experiment to realize the MRR-KLM-CNOT.

The MRR based architecture features an enhanced parameter space in comparison with bulk optical versions for any given quantum optical network. We first demonstrated this theoretically with respect to the Hong-Ou-Mandel (HOM) Effect  \cite{HOM:1987} in a double bus microring resonator (db-MRR).
Specifically, we showed that the topology of the db-MRR produces Passive Quantum Optical Feedback (PQOF) in the transition amplitudes available to the two-photon state vector resulting in a dimensional dilation of the success manifolds for the HOM from a zero-dimensional, single operating point, viz. a 50/50 beam splitter, to multi-dimensional manifolds described by continuously variable values of the physical coupling parameters that satisfy the HOM constraint. Success in the HOM context being the production of the two-photon NOON state    
$\ket{2,0}+e^{i\,\varphi}\,\ket{0,2}$,
at the output \cite{Kok:2002, Gerry_Knight:2004}. A projective measurement on either output mode of a system in the state   will result in the detection either of both photons, or of none. Correlations counts on such measurements will reveal the absence of photon coincidences, owing to the absence of the   branch of the output two-photon state vector. This measurement induced nonlinearity distinguishes the HOM Effect as an important tool for quantum optical information processing systems. The inherent scalability of the db-MRR and the tunability of the HOM Effect owing to the existence of the HOM Manifolds (HOMM) are the main motivations behind our identification of the db-MRR as a fundamental circuit element for scalable, linear quantum optical networks for information processing \cite{Hach:2014}.
The CNOT in any architecture relies upon the function of two NonLinear Phase Shift Gates NLPSGs), each of which is a quantum circuit of non-trivial complexity. The basic role of the serially connected NLPSGs within the KLM CNOT is to remove, via destructive interference, any two-photon states resulting from the Hong-Ou-Mandel-like ‘bunching’ of control and target photons within the device. The KLM CNOT results in the correct truth table if and only if exactly one photon serves as the control qubit and exactly one other photon serves as the target qubit. The device will fail for any instance in which both of the input photons emerge from the same output. Though the detailed network geometries differ slightly, this is a hard requirement for the success of the KLM CNOT based on both polarization and dual-rail encoded photonic qubits. 
Each NLPSG performs the local isometry
\be{NLPSG:eqn}
\a_0\ket{0} + \a_1\ket{1} + \a_2\ket{2} \xrightarrow{\text{NLPSG}}  \a_0\ket{0} + \a_1\ket{1} - \a_2\ket{2}, 
\ee
on the target mode, conditioned on the outcome of a projective measurement performed on two ancillary modes \cite{KLM:2001, Skaar:2004}. This measurement induced nonlinear phase shift is the essence of the NLPSG. 
In this paper, we propose a direct interferometric test of the NLPSG. To our knowledge, this is the first proposed test for the isolated NLPSG. While the demonstration of the KLM CNOT is a seminal proof-of-concept experiment that indirectly verifies the function of the bulk optical NLPSGs involved, it does not provide a diagnostic result for an individual NLPSG, and therefore, it is not portable to investigations of other potential applications for the NLPSG. 
Advances in integrated photonics have vastly expanded experimental and even manufacturing capabilities for the design and implementation of linear quantum optical networks. Unlike the situation with bulk optics, scalable quantum circuits are ever becoming commonplace in silicon nanophotonics. The prior hurdle to increase circuit complexity was the waveguide propagation loss, but through combined efforts of researchers and foundries that loss has been reduced to levels which allows quantum integrated photonics to flourish. The ability to have access to integrated photonics foundries has been game changing for the field, allowing scalable and reproducible quantum devices to be fabricated beyond scales physically impossible through any other means.  Thus more complex circuit designs can be constructed, expanding the parameter space which we have access to manipulate in these circuits, such as the construction of higher fidelity quantum gates.    
A robust and experimentally feasible means of performing quality assurance tests on the essential component for a KLM CNOT that until now has simply been an experimental “black box.” By considering photon triple-coincidence counts on the success mode (i.e. Mode 2) of the NLPSG along with the output modes of the Mach Zehnder Interferometer (MZI) shown in Fig. (1), we show that a signature of the successful operation of the NLPSG can be measured with sufficient visibility even in the experimentally relevant case of lossy click/no-click detections with weak coherent states in modes 1 and 4. Further, based on a suggestion from Professor Paul Kwiat 
\footnote{We graciously acknowledge Paul Kwiat for this insightful suggestion conveyed to us at the 1st Photons for Quantum (PfQ) Conference, Rochester Institute of Technology, Rochester, NY, 23-25Jan2019}, 	
we demonstrate that our direct test is significantly enhanced by the use of output from colinear Spontaneous Parametric Down Conversion (cl-SPDC) in each arm of the MZI.

The outline of this paper is a follows. 
In \Sec{sec:KLM:NLPSG} we review the operation of the KLM NLPSG on three modes (one primary mode, and two ancilla modes), and describe our MZI setup for the direct test of the KLM and MRR NLPSG.
In \Sec{sec:NLPSG:under:U} we derive the conditions for successful operation of the NLPSG under the action of an arbitrary unitary transformation. 
Before we embark on the calculation for the coincidence interference probability,  we first  derive in \Sec{sec:pncc:pnc} the POVM for non-photon number resolving \tit{click/no-click} detection with finite detection efficiencies typical of many laboratory experiments. 
In \Sec{sec:direct:meas:NLPSG} we begin our main calculation, and derive the primary interference effect of the coincidence probability using an MZI setup with a NLPSG in one leg and a phase shifter in the other leg. We derive the various interference and accidental output states generated by the even and odd number photon states of weak coherent state (w-CS) inputs, 
containing up to two photons, in each arm of the MZI.
We examine effect of colinear spontaneous parametric down conversion (cl-SPDC) input states that do not contain the single photon branch, and see that they generate the significant portion of the coincidence interference effect generated by w-CS.
Finally, in \Sec{sec:Conclusion} we conclude, and discuss the significance of this work for photonic  integrated waveguide devices.
In the appendices we review the essentials of the KLM and the MRR NLPSG and their maximum success probabilities.
Additionally, we remind the reader of the action of a BS on a product of photon Fock states at its inputs ports, 
which will be needed for the MZI calculation in \Sec{sec:direct:meas:NLPSG}. Finally, in the last appendix  we explicitly list the coefficients of the four and five photon accidental states that are generated along with the primary coincidence interference effect.

\section{The NLPSG}\label{sec:KLM:NLPSG}
As discussed in \Sec{sec:Intro}, the KLM NLPSG imparts a phase shift of $\pi$ on the two-photon branch of any single-mode-1 (normalized) state that evolves through it,
\bea{NLPSG:effect}
\ket{\psi_{(in)}}_{123} &=&  (\a_0\ket{0}_1 + \a_1\,\ket{1}_1 + \a_2\ket{2}_1)\otimes \ket{1,0}_{2,3},  \no
\xrightarrow{NLPSG} \ket{\psi_{(out)}}_{123} &=& (\a_0\ket{0}_1 +\a_1\,\ket{1}_1 - \a_2\ket{2}_1)\otimes \ket{1,0}_{2,3}, 
\eea
with with $|\a_0|^2+|\a_1|^2+|\a_1|^2=1$. Typically, this state will be generated as a weak coherent state (w-CS) 
with mean number of photons $\bar{n}_1=\alpha^2\ll1$, where $\a_k = e^{-|\a|^2/2}\frac{\a^k}{\sqrt{k!}}$. While the input state in mode-1 can be of a general form containing up to two photons, for simplicity we will refer to it in this work as a w-CS.
Currently, there is no known way to affect the transformation in \Eq{NLPSG:effect} deterministically and nondestructively
via unitary evolution. Instead, the transformation is realized probabilistically by using two auxiliary optical modes, here labeled 2 and 3, with
one input photon in ancilla mode 2. 
Projecting out the final state conditioned on a \tit{click} on mode-2 and \tit{no-click} on mode-3 
produces the desired local isometry on mode-1 \Eq{NLPSG:effect}.
It has been shown \cite{Skaar:2004,Obrien:2011} that this action is successful with a maximum
probability of 1/4 , and that the result of the projective measurement faithfully indicates the success of the transformation.
Consequently, the optimal probability of success for the KLM or MRR
CNOT gate is 1/16 [1,4], which employs two NLPSG. 
This NLPSG-based CNOT gate effectively performs a HOM \cite{HOM:1987,Alsing_Hach:2019} interference on the two-photon branch 
$\ket{2}_1$ of mode-1, in order to affect the CNOT operation on the remaining branch of mode-1, $\a_0\ket{0}_1 +\a_1\,\ket{1}_1$.

In this work we consider a direct interferometric coincidence detection of the success probability for both the KLM and MRR implementations of the NLPSG through their insertion into one (upper) leg (mode $1$, with ancilla modes $2, 3$) of a Mach Zehnder interferometer (MZI) 
\begin{figure}[ht]
\includegraphics[width=4.25in,height=2.55in]{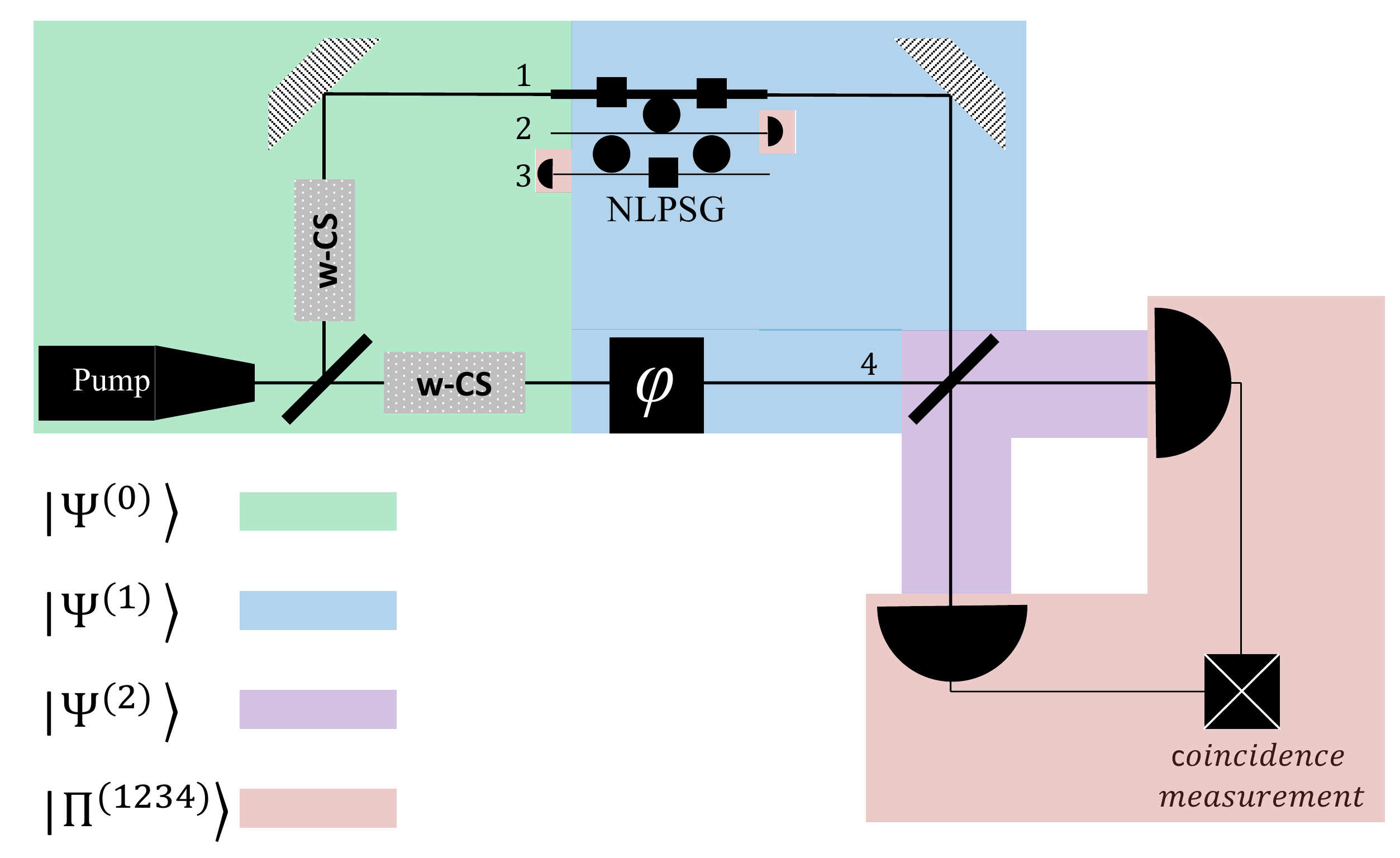}
\caption{ (Color online) 
The MRR NLPSG, with the ordinary three beam splitters used in the KLM implementation (see \Fig{fig:skaar:nlsg} in \App{app:KLM:MRR:NLPSG})
 replaced by micro-ring resonators (MRRs) (black circles). 
The input state $\ket{\Psi^{(0)}}$ entering the upper and lower leg of the MZI are  weak coherent states (w-CS) of the form
$(\a_0\ket{0}_1 +\a_1\ket{1}_1+ \a_2\ket{2}_1)$  and
$(\a'_0\ket{0}_4 +\a'_1\ket{1}_4+ \a'_2\ket{2}_4)$, respectively,
with the modes of the NLPSG in the upper leg of the MZI labeled (top-down) as $1,2,3$.
Modes $2$ and $3$ are the ancilla modes to the NLPSG that are initially in the state $\ket{1,0}_{2,3}$.
The lower leg, mode 4,  of the MZI contains a PHASE shift element $e^{ i  \varphi a_4^\dag a_4}$ which effectively sends $\a'_k\to\a'_k\,e^{i\,k\,\varphi}$.
The action of $NLPSG_{123}\otimes PHASE_4$ produces the intermediate state $\ket{\Psi^{(1)}}$.
The final element of the MZI is a BS of angle $\theta$ (such that $\theta=\pi/2$ is a 50:50 BS), whose output is $\ket{\Psi^{(2)}}$.
Coincidence detection, producing the un-normalized state $\ket{\tilde{\Psi}^{(2)}} = \Pi^{1234}\ket{\Psi^{(2)}}$,
 is performed on the exiting modes 1 and 4, conditioned on the NLPSG ancilla modes $2$ and $3$, occurs with
 probability $\IP{\tilde{\Psi}^{(2)}}{\tilde{\Psi}^{(2)}}$. 
The KLM NLPSG is obtained by replacing each black circle in the NLPSG sub-diagram 
by a single beam splitter with reflectivities $0\le\eta_i \equiv r_i^2\le 1$, where 
and  $-1\le r_i\le 1$ are reflection coefficients for $i\in\{1,2,3\}$.
}\label{fig:direct:meas:nlpsg}
\end{figure}
and a PHASE shift element in the other (lower) leg (mode $4$), as shown in \Fig{fig:direct:meas:nlpsg}.
We will consider the case of finite detection efficiencies $\xi_k<1$ in each mode $1,2,3,4$, which can also can be considered as incorporating propagation and scattering losses. This will allow us to measure the success probability of the NLPSG in the presence of \tit{accidentals}, namely, those coincidence counts that arise from states that are outside the isometry in \Eq{NLPSG:effect}. 
These accidentals add a noise floor to the the primary interference effect upon output from the MZI due to the mixing action of the BS and the use of detectors with finite detection efficiencies.

The initial state of the  system entering the MZI is 
\be{Psi:0}
\hspace{-.6in}
\ket{\Psi^{(0)}} =  
 (\a_0\ket{0}_1 + \a_1\,\ket{1}_1 + \a_2\ket{2}_1)\otimes \ket{1,0}_{2,3} \otimes 
 (\a'_0\ket{0}_4 + \a'_1\ket{1}_4 + \a'_2\ket{2}_4), \quad
 \sum_{k=0}^{2} |\a_k|^2 =  \sum_{k=0}^{2} |\a'_k|^2 = 1,
\ee
with modes $1, 2, 3$ associated with the NLPSG in the upper leg of the MZI and mode 4 in the lower leg 
(see labeling scheme in  \Fig{fig:direct:meas:nlpsg}).
We can intuitively understand why this state will produce an interference pattern upon coincidence detection of modes 1 and 4 exiting the MZI.
The lower leg of the MZI contains a PHASE shift element $e^{ i  \varphi a_4^\dag a_4}$ which effectively sends $\a'_k\to\a'_k\,e^{i\,k\,\varphi}$.
Recall that a lossless unitary BS preserves the total photon number entering its ports. 
Thus, as discussed in detail in \App{app:BS}, a state $\ket{n}_a\ket{m}_b$ entering a BS will generate the $n+m+1$ states $\{\ket{p}_a\ket{n+m-p}_b\}$ where $p\in{0,1,\ldots,n+m}$ with (Wigner) rotation coefficients amplitudes. 
Thus, upon exit from the MZI, the input states $\ket{0}_1\ket{2}_4$ and   $\ket{2}_1\ket{0}_4$ to the final BS will generate the 
output state $\ket{1}_1\ket{1}_4$ with phase factors proportional to $e^{i\,2\, \varphi}$ and $1$ respectively, and BS-dependent modified amplitudes.
Similarly, the input state $\ket{1}_1\ket{1}_4$ to the final BS will also generate the output state  $\ket{1}_1\ket{1}_4$ 
with phase factor $e^{i\,\varphi}$ and a BS-dependent modified amplitude.
These two sets of terms, which we will consider individually, contribute to the primary coincidence interference pattern when we 
condition on the \tit{click}/\tit{no click} of the ancilla modes 2 and 3. Here a  \tit{click} detection means that, sans photon number resolving detectors, typical laboratory photon counting experiments are performed with \tit{bucket detectors} (e.g. average efficiency APDs with $\xi_k\simeq 40\%$, or high efficiency SNSPDs  with $\xi_k\simeq 85\%$)  with the probability to detect $n$ photons scaling as $\xi^n$. Such higher order detections are called \tit{accidentals} and contribute an additional noise floor (over that of detector dark counts, which we assume for simplicity to be zero) to the coincidence measurements. 

In this work, we will keep track of such accidental terms using 
the reasonable approximation of detecting at most two photons in any single mode, $1, 2, 3, 4$.
We will see that upon output the set of input states  $\{ \ket{0}_1\ket{2}_4, \ket{2}_1\ket{0}_4\}$ will generate the output state 
$\ket{1,1,0,1}_{1234} = \ket{1,1}_{14}\ket{1,0}_{23}$ upon which the primary interference will be observed, 
with the output ancilla modes remaining in their ideal ``success heralding" state $\ket{1,0}_{23}$.
The output state will also contain (orthogonal) 5-photon states with the the output ancilla modes not necessarily remaining in $\ket{1,0}_{23}$, 
plus various other photon Fock states in modes 1 and 4. 
Similarly, upon output, the input state $\ket{1,1}_{14}$ will also generate the ideal  success state $\ket{1,1,0,1}_{1234}$, 
as well as 4-photon accidental states.
A little forethought indicates that the amplitude of the output state $\ket{1,1,0,1}_{1234}$ will be of the form
$(\a_0\,\a'_2\,A\,e^{i\,2\,\varphi } + \a_2\,\a'_0\,B +\a_1\,\a'_1\,C e^{i \,\varphi })$
$\approx \a\,e^{i \,\varphi} \,( A(\theta)\,e^{i \,\varphi} + B(\theta)\,e^{-i \,\varphi} +C(\theta))$ 
leading to a primary success probability (squared amplitude) varying as
$DC(\theta) + 2\, A(\theta)\,B(\theta)\,\cos(2\,\varphi) + 2\,C(\theta)\,\big(A(\theta)+B(\theta)\big)\cos(\varphi)$. Here $A(\theta), B(\theta) ,C(\theta)$ (taken real for simplicity) will depend on the final BS angle $\theta$, and $DC(\theta) = A(\theta)^2+B(\theta)^2+C(\theta)^2$ is the constant (independent of $\varphi$) contribution assuming unit detection efficiencies. When finite detection efficiencies are taken into account, there will be a prefactor scaling as $\xi^6\ll 1$ (assuming, for simplicity,  equal detection efficiencies in all modes ) 
as well as both an ``$AC(\theta,\phi)$" accidental term (dependent on the phase angle $\varphi$,  
arising from 4-photon output states generated from 
input states containing $\ket{1}_1$,\, $\ket{1}_4$, or both) and 
a $DC(\theta)$  accidental term
(arising from the 5-photon states generated by the $\ket{0,2}_{1,4}$ and $\ket{2,0}_{1,4}$ input states).
Both these accidental states will contribute to the measured coincidence counts. 
However, these terms will be down in magnitude by factors of $\bar{n}\ll 1$ and $\bar{n}^2\lll1$ respectively, from the primary interference probability.
The details supporting this intuition will be worked out explicitly in the following sections.

\section{The NLPSG under arbitrary unitary evolution}\label{sec:NLPSG:under:U}
Before we begin the main coincidence measurement calculation, let us first demonstrate the action of the unitary operator $U$ representing the NLPSG on modes 1, 2, 3.
Under an arbitrary $N\times N$ unitary evolution $U$ on mode-$j$, the boson creation operators $a^\dag_{in}$ 
are transformed linearly via \cite{Skaar:2004}
\be{U:S}
a_{j, in}^\dag\to U\,a_{j, in}^\dag\,U^\dag = \sum_{k=1}^{N} S^{T}_{jk}\,a_{k, out}^\dag =   \sum_{k=1}^{N} a_{k, out}^\dag \,S_{kj}, 
\ee
where $T$ is transpose,
defining the corresponding unitary matrix of coefficients $S_{kj}$ that act as transition coefficients for a photon initially in mode $j$ to be routed to output mode $k$. Henceforth, we will drop the $in, out$ subscript labels.
The action of $U$ on the input state
$\ket{\psi_{(in)}}_{123} =  (\a_0\ket{0}_1 + \a_1\,\ket{1}_1 + \a_2\ket{2}_1)\otimes \ket{1,0}_{2,3}$ of \Eq{NLPSG:effect} 
is then given by \cite{Alsing_Hach:2019}
\bsub
\bea{Psi:0:123}
\ket{\psi_{(in)}}_{123} &=&  (\a_0\ket{0}_1 + \a_1\,\ket{1}_1+  \a_2\ket{2}_1)\otimes \ket{1,0}_{2,3}, \\
&=& \left(\a_0 + \a_1 \, a_1^\dag + \a_2\,\frac{a_1^{\dag 2}}{\sqrt{2}}\, \right)\,  a_2^\dag \,\ket{0,0,0}_{123}, \no
%
\xrightarrow{S}  \ket{\psi_{(out)}}_{123} 
&=&
\left[
\a_0 + \a_1 \, \sum_{j=1}^3 S_{j 1} a^\dag_{j} +  \a_2\,\frac{1}{\sqrt{2}}\,\left( \sum_{j=1}^3 S_{j 1} a^\dag_{j}\right)\,\left( \sum_{k=1}^3 S_{k 1} a^\dag_{k}\right)\,
\right]\,
\left( \sum_{\ell=1}^3 S_{\ell\,2} a^\dag_{\ell} \right)\,
\ket{0,0,0}_{123}, \qquad \no
&\equiv& \ket{\psi^{NLPSG}_{(out)}}_{123} + \ket{\psi^{\perp}_{(out)}}_{123},  \label{NLPGS:perp:out:states:defn} 
%
\eea 
\esub
where we have defined the 3-photon NLPSG state $ \ket{\psi^{NLPSG}_{(out)}}_{123}$ as
\be{NLPSG:defn}
\ket{\psi^{NLPSG}_{(out)}}_{123}
\equiv
\big(
\beta_0\, \a_0\,\ket{0}_1 + \beta_1\,\a_1\,\ket{1}_1 - \beta_2\,\a_2\,\ket{2}_1
\big) 
\otimes \ket{1,0}_{2,3} 
\ee
with the $\beta_k$ coefficients defined as 
\bsub
\bea{betas:defn}
\textrm{Condition-0:} \quad  \beta_0 &=&  S_{22}, \label{beta:0}\\
\textrm{Condition-1:} \quad  \beta_1 &=&  S_{11}\,S_{22} + S_{21}\,S_{12}, \label{beta:1}\\
\textrm{Condition-2:} \quad \beta_2 &=&  -S_{11}\, \left( S_{11}\,S_{22} + 2\, S_{21}\,S_{12} \right), \label{beta:2}
\eea
\esub
and 
\be{psi:perp:w-CS}
\hspace*{-.25in}
\ket{\psi^\perp_{out}}_{123} \equiv
\left[
\a_0 \, \sum_{\ell\ne2} S_{\ell\,2} a^\dag_{\ell}  + 
\a_1 \, \hspace{-0.25in}\sum_{j,\ell\ne\{(1,2),(2,1)\}} 
 \hspace*{-.3in} S_{j 1} S_{\ell\,2} a^\dag_{j} a^\dag_{\ell} +
\frac{1}{\sqrt{2}}\;\a_2
\hspace*{-.15in} \, \sum_{j,k,\ell\ne\{\trm{perm}(1,1,2)\}}
 \hspace*{-.35in} S_{j 1} S_{k 1} S_{\ell\,2}\, a^\dag_{j} a^\dag_{k} a^\dag_{\ell} 
\right]\,
\ket{0,0,0}_{123}, 
\ee 
as the remaining ``non-NLPSG"  state orthogonal to $\ket{\psi^{NLPSG}_{(out)}}_{123}$.
Successful operation of the NLPSG occurs when all three conditions \Eq{beta:0}, \Eq{beta:1}, and \Eq{beta:2} hold simultaneously, namely
$\beta_0=\beta_1= \beta_2 \equiv \beta$, in which case  
$\ket{\psi^{NLPSG}_{(out)}}_{123} \to \beta\, (a_0\,\ket{0}_1 + a_1\,\ket{1}_1 - \a_2\,\ket{2}_1$ with success probability $|\beta|^2$.
The self consistency of all three conditions requires $S_{11} = 1\mp\sqrt{2}$, with the physical solution ($|S_{11}|\le1$) demanding the solution with the minus sign. The remaining two conditions then demand that 
\be{NLPSG:success:prob}
P^{NLPSG}_{\trm{success}} = |\beta|^2 = |S_{22}|^2 = \frac{1}{2}\,|S_{21}|^2\,|S_{12}|^2, \qquad S_{11} = 1-\sqrt{2}.
\ee
This is the operational scenario for the use of two NLPSG in the KLM-CNOT  gate \cite{Obrien:2011,Alsing_Hach:2019,Alsing_Hach:2018}. At this stage, the unitary transformation $S$ is arbitrary. In the case of the KLM NLPSG implementation, $S$ is the product of three BS operators. For the MRR NLPSG implementation, as explored in \cite{Alsing_Hach:2019} and discussed in \App{app:KLM:MRR:NLPSG}, $S$ is the product of three MRR transfer matrix operators. Both these cases will be explored below, but for now we can remain unitarily agnostic, with a general $S$ matrix.

Finally, we note that if one's sole purpose is simply to test the successful  sign flip on the state $\ket{2}_1$, (say as an alternative to testing of the validity of the NLPSG with w-CS inputs) then this could also be accomplished by setting $\a_1\equiv 0$, using a co-linear SPDC  (cl-SPDC) input state $\a_0\,\ket{0}_1 + \a_2\,\ket{2}_1$ 
 and lastly, only requiring that 
Condition-0 \Eq{beta:0}, and Condition-2 \Eq{beta:2} hold, namely $\beta_0=\beta_2\equiv \beta$, with the value of $\beta_1$ unconstrained.
While it is easier to generate w-CS than cl-SPDC states, the former which are also more operationally useful in optical quantum computing scenarios, it is informative to also explore the details of the latter case. 
It will turn out that both types of input states produce nearly identical coincidence interference patterns when the cl-SPDC input state scenario employ detectors operating at $40\%$ detection efficiencies, and the w-CS input state scenario employ detectors with $85\%$ detection efficiencies, 
both with NLPSG success probabilities of $|\beta|^2=1/4$. 
We will discuss the cl-SPDC scenario in \Sec{sec:direct:meas:NLPSG}. 
For now we will explore the case of the general w-CS input state \Eq{Psi:0:123}. 

Before we begin the analysis of the MZI interferometer with a NLPSG in one leg and a PHASE shifter in the other, 
we first examine the POVM operator that is needed to project out the final state (from the MZI-transformed pure input state) that contributes to the coincidence counts.

\section{\tit{click} and \tit{no-click} detection projection operators}\label{sec:pncc:pnc}
Since the NLPSG is realized non-deterministically, we first review the concept of non-photon number resolving detection 
(bucket or \tit{click/no-click} detection) that is typical of many laboratory experiments.

\subsection{Single mode detection}
Consider a detector with probability (detection efficiency) $0\le\xi\le1$ to detect one photon in a single mode $\ket{1}$, with the corresponding probability $1-\xi$  not to detect the single photon. 
Then the projection operators $\Pi_{NC}$ and $\Pi _{C}$ for a \tit{no-click} and a \tit{click} detection, respectively
 (i.e. non-photon number resolving detection) are given by
\bsub 
\bea{PNCC:PNC:PC}
\Pi_{NC} &=& \sum_{n=0}^{\infty} \left( 1-\xi \right)^{n} |n \rangle \langle n | \to \ket{0}\bra{0} \;\; \textrm{as} \;\; \xi\to 1,\label{eqn:PNC} \\
\Pi_{C} &=& I - \Pi_{NC} = \sum_{n=0}^{\infty} \left[ 1 - \left( 1-\xi \right)^{n} \right] |n \rangle \langle n | 
\to  \sum_{n=1}^{\infty} \ket{n}\bra{n}  = I -\ket{0}\bra{0}  \;\; \textrm{as} \;\; \xi\to 1, \label{eqn:PC} 
\eea
\esub
and  hence the pair 
\be{1mode:POVM}
\textrm{Single-mode detection POVM} = \{\Pi_{C},  \Pi_{NC}\equiv I - P_{CC}\},
\ee
forms a dichotomous single mode detection POVM.
Here, \Eq{eqn:PNC} is intuitively understood as the probability $(1-\xi)^n$ not to detect the state $\ket{n}$ of $n$ photons, 
and for the \tit{no-click} projector $\Pi_{NC}$ we then sum over all possible photon number states. In the limit of perfect (photon number resolving) detection $\xi\to1$, we have that $\Pi_{NC}$ is just the projection onto the vacuum state $\ket{0}\bra{0}$. The opposite case of the detection one or more photons (a \tit{click}) in the given mode is  trivially given as $I-\Pi_{NC}$, 
with the intuitive $\xi\to1$ limit of  $I -\ket{0}\bra{0} = \ket{1}\bra{1} +  \ket{2}\bra{2} + \cdots$ 
(i.e. the projector onto the state containing one or more photons).

The unnormalized state $\ket{\tilde{\Psi}'}$ just after a \tit{click} detection event is given by 
$\ket{\tilde{\Psi}'} = \Pi_C\,\ket{\Psi}$ for the pure state $\ket{\Psi}$ just before the measurement. 
(Note: throughout the paper, we used a \emph{tilde} to indicate an unnormalized state, whose norm yields a probability). 
The probability for the \tit{click} measurement is then just the norm of this state $P_C = || \ket{\tilde{\Psi}'}||^2 =  \bra{\Psi} \Pi^2_{C} \ket{\Psi}$.
(Note that while $\Pi_C$ is a measurement projection operator, it is not a \tit{von-Neumann projection} operator 
in the sense that $\Pi^2_{C}\ne \Pi_{C}$. 
Along with $\Pi_{NC}$, it is an element of a POVM). This gives the expressions
\bsub
\bea{PNCC:pNC:pC}
\hspace{-1in}
P_{NC} &=& \textrm{Tr}  [\Pi_{NC}\, \ket{\Psi}\bra{\Psi}] =
 \sum_{n=0}^\infty\, q^2_n\, |\bra{n} \Psi\rangle|^2, \hspace{0.75in}  q_n = (1-\xi)^n, \;\; q_0 = 1, \; q_1 = 1-\xi, \label{pNC}\\
P_{C} &=&  \textrm{Tr}  [\Pi_{C}\, \ket{\Psi}\bra{\Psi}] = 
\sum_{n=0}^\infty\, p^2_n \,|\bra{n} \Psi\rangle|^2= 1-p_{NC},  \;\; p_n = 1-q_n, \quad\;\;  p_0=0, \;\; p_1=\xi. \label{pC}
\eea
\esub
\subsection{Many mode detection}
We can easily extend the concept of \tit{click} and \tit{no-click} detection to many modes.  
Consider first two modes $a$ and $b$. If one had \textit{perfect} detection efficiency $\xi\to1$, the situation in which we 
\textit{do not} have a simultaneous coincidence click between  modes $a$ and $b$ is given by
\be{pncc:eta:1:reduction}
\Pi_{NCC} \myover{\rightarrow}{\xi\rightarrow 1} \ket{0}_a\bra{0}\otimes I_b +  I_a \otimes\ket{0}_b\bra{0} - \ket{0}_a\bra{0}\otimes \ket{0}_b\bra{0},
\ee
where the first term is "no-click" in detector $A$ and anything in detector $B$, the second term is the reverse situation, and
the last term with the "-" sign is needed to avoid the double counting of the vacuum projection $\ket{0}_a\bra{0}\otimes \ket{0}_b\bra{0}$ that occurs in the first two terms.

To extend \Eq{pncc:eta:1:reduction} to imperfect detection  $0\le \xi_a, \xi_b\le 1$, we utilize \Eq{eqn:PC} to extend 
$I_a~-~\ket{0}_a\bra{0} \rightarrow  \Pi^{(a)}_{C} = \sum_{n=0}^{\infty} 
\left[ 1 - \left( 1-\xi_a \right)^{n} \right]\,\ket{n}_a\bra{n} \equiv \sum_{n=0}^{\infty} p^{(a)}_{n} \ket{n}_a\bra{n}$ with 
$p^{(a)}_n=[1 - \left( 1-\xi_a \right)^{n}]$ the probability \textit{to detect} a ``click'' of $n$ photons in mode-$a$ Fock state $\ket{n}_a$. 
Note that $p^{(a)}_0 =0$ and $p^{(a)}_1 = \xi_a$.
Then, the probability to detect a click in both mode-$a$ and in mode-$b$, i.e. a \textit{coincidence count} (CC),  with finite detection efficiencies is just the \textit{product} of the individual probabilities for mode-$a$ and mode-$b$, 
corresponding to the product of the projection operators for each mode,
namely
\bsub
\bea{PCC}
\Pi^{(a b)}_{CC} =  \Pi^{(a)}_{C}\otimes \Pi^{(b)}_{C} &=& \sum_{n=0}^{\infty} p^{(a)}_n \ket{n}_a\bra{n} \otimes \sum_{m=0}^{\infty} p^{(b)}_m \ket{m}_b\bra{m}, \\ 
&\equiv& \sum_{n=0}^{\infty}\,\sum_{m=0}^{\infty} p^{(ab)}_{n m}  \ket{n,m}_{ab}\bra{n,m} , \quad 
        p^{(ab)}_{n m} = [1 - \left( 1-\xi_a \right)^{n}]\,[1 - \left( 1-\xi_b \right)^{m}]. \quad
\eea
\esub
We see that the above expression has the correct limits, namely $p^{(ab)}_{00} =p^{(ab)}_{n 0}=p^{(ab)}_{0m}=0$ appropriate for \textit{not} detecting a coincidence click, and   $p^{(ab)}_{11} = \eta_a\,\eta_b$.
Lastly, the above expression reduces in the limit of unit detection efficiencies to 
$[I_a-\ket{0}_a\bra{0}]\otimes[I_b-\ket{0}_b\bra{0}]$ such that in the same limit the probability for \textit{no coincidence counts} (NCC) 
$\Pi_{NCC}=I_a\otimes I_b - \Pi_{CC}$ reduces to the correct limiting form given by \Eq{pncc:eta:1:reduction}.
Thus, the   dichotomous two-element POVM defining two-mode coincidence \textit{click/no-click} detection is given by
\be{2mode:POVM}
\textrm{Two-mode detection POVM}_{a,b} = \{\Pi^{(ab)}_{CC},  \; \Pi^{(ab)}_{NCC}\equiv I_a\otimes I_b - \Pi^{(ab)}_{CC}\}.
\ee
This is easily generalized to arbitrary simultaneous coincidence clicks on  $M$ modes $a_{i\in\{1,2,\ldots,M\}}$  via
\bsub
\bea{Mmode:POVM}
\textrm{$M$-mode detection POVM}_{a_{1},\ldots,a_{M}} 
&=& \{\Pi^{({a_{1},\ldots,a_{M}})}_{CC},  \; P^{({a_{1},\ldots,a_{M}})}_{NCC}\equiv 
I_{a_1}\otimes\ldots\otimes I_{a_M} - \Pi^{({a_{1},\ldots,a_{M}})}_{CC}\}, \qquad \\
\Pi^{(a_{1},\ldots,a_{M})}_{CC} &=& \bigotimes_{i=1}^M  \Pi^{(a_i)}_{C }, \quad 
p^{(a_{1},\ldots,a_{M})}_{n_1\ldots\,n_M} = \prod_{i=1}^M p^{(a_i)}_{n_i} =  \prod_{i=1}^M [1 - \left( 1-\xi_{a_i} \right)^{n_i}].
\eea
\esub

The takeaway point of this section is as follows. Under perfect detection efficiency, $\xi_i=1$ only the state 
$\ket{1,1,0,1}_{1234}$ will contribute to the probability interference pattern, as discussed in \Sec{sec:KLM:NLPSG}.
However, under finite, imperfect detection efficiencies, $\xi_i<1$,  output states other than  $\ket{1,1,0,1}_{1234}$ will also contribute to the output detected signal with varying probabilities. We will call such states \tit{accidentals}, since they arise due to finite detection efficiencies.  Note that in order to contribute to the total output signal, such states \tit{must} contain at least one photon in each of modes 1, 2, and 4, and any number of  photons in mode 3, i.e.  $\ket{n_1, n_2, n_3, n_4}_{1234}$ with $n_1, n_2, n_4 \ge 1, \; n_3\ge 0$.
\section{Direct measurement of the NLPSG}\label{sec:direct:meas:NLPSG}
In this section we analyze the MZI given in \Fig{fig:direct:meas:nlpsg} containing the NLPSG in the upper leg of the MZI, with primary mode 1 and ancilla modes 2 and 3,  and the PHASE shift element in the lower leg, mode 4. After the action of 
$NLPSG_{1,2,3}\otimes PHASE_4$, modes 1 and 4 interfere on a $BS_{14}$, and are subsequently coincidently detected upon exit from the MZI, while we simultaneously ask for a \tit{click} detection on mode 2 and a \tit{no-click} detection on mode 3.
Our unitary operator is given by $\U = BS_{14}\cdot(NLPSG_{123}\otimes PHASE_4)$ and our projection operator will be
$\Pi^{(1,2,3,4)} \equiv \Pi^{(1)}_{C}\otimes \Pi^{(2)}_{C}\otimes \Pi^{(3)}_{NC}\otimes \Pi^{(4)}_{C}$.
Note that we will explicitly implement (by hand) the phase shift element $PHASE_4 = e^{i\,\varphi\,a^\dag_4\,a_4}$ on mode 4, which simply has the net effect to transforming $\a'_k\to\a'_k\,e^{i\,k\,\varphi}$ on the w-CS$_4$ input state.

\subsection{Preliminaries}
As before, we allow the KLM triple BS (or triple MRR) operator on modes $1,2,3$ to be represented by $S_{ij}$, and the BS
 transformation on modes $1,4$ to be represented by $B_{ij}$.
Extending these operators to $4\times 4$ matrix representations, we define
\be{S:B:4x4}
\B = 
\left[
\begin{array}{cccc}
\cos(\theta/2)  & 0 & 0 &  \sin(\theta/2) \\
0                     & 1 & 0 &       0 \\
0                     & 0 & 1 &       0 \\
-\sin(\theta/2) & 0 & 0 & \cos(\theta/2)\end{array}
\right],\; \qquad
\S = 
\left[
\begin{array}{cccc}
S_{11} & S_{12}  & S_{13}  & 0  \\
S_{21} & S_{22}  & S_{23}  & 0  \\
S_{31} & S_{32}  & S_{32}  & 0  \\
0         & 0           &  0          & 1\end{array}
\right], \qquad
%
%
\ee
where the rows and columns are labeled by the mode indices in the order $\{1,2,3,4\}$.
(Note, the choice of the argument $\theta/2$ in the BS is so that a 50:50 BS is given by $\theta=\pi/2$).
We define the product of these matrices as the unitary $\U$ 
\be{U}
\U \equiv \B\,\S = 
\left[
\begin{array}{cccc}
\cos(\theta/2)\,S_{11} & \cos(\theta/2)\,S_{12}  & \cos(\theta/2)\,S_{13}  & \sin(\theta/2)  \\
S_{21} & S_{22}  & S_{23}  & 0  \\
S_{31} & S_{32}  & S_{32}  & 0  \\
-\sin(\theta/2)\,S_{11} & -\sin(\theta/2)\,S_{12}  & -\sin(\theta/2)\,S_{13}  & \cos(\theta/2) 
\end{array}
\right].
\ee
The unitary transformation $\U$  affects the following transformations on the boson creation operators
\be{S:to:B:to:U}
a^\dag_i \xrightarrow{S} \sum_{j=1}^4\, a^\dag_j\,\S_{ji} 
               \xrightarrow{BS} \sum_{j=1}^4\,  \sum_{k=1}^4\, a^\dag_k \,\B_{kj} \,\S_{ji} 
               \equiv \sum_{k=1}^4\, a^\dag_k\,\U_{ki}, \quad 
              \textrm{with}\;\;  \U_{ki} = \sum_{j=1}^4\, \B_{kj} \,\S_{ji}.
\ee
This allows us to transform the initial state $\ket{\Psi^{(0)}}_{1234} \xrightarrow{\U}\ket{\Psi^{(2)}}_{1234}$ (see \Fig{fig:direct:meas:nlpsg}).
Upon coincidence detection of modes 1 and 4, with \tit{click/no-click} detection on modes 2 and 3, we have the unnormalized post-measurement state $\ket{\Psi^{(1)}}_{1234} \xrightarrow{\Pi^{(1234)}}  \ket{\tilde{\Psi}^{(2)}}_{1234}$ (indicated with a \emph{tilde}) with detection probability
$P_{1234} = ||  \ket{\tilde{\Psi}^{(2)}}_{1234} ||^2$.


We begin by writing the initial state $\ket{\Psi^{(0)}}_{1234}$ ,  using 
$\sum_{i=0}^2 |\a_i|^2 =  \sum_{i=0}^2 |\a'_i|^2 =1$
as
\bsub
\bea{Psi:0:w-CS}
\hspace{-1in}
\ket{\Psi^{(0)}} _{1234}
&=&  \big[\a_0\ket{0}_1 +a_1\ket{1}_1 +\a_2\ket{2}_1\big]\otimes \ket{1,0}_{2,3} \otimes
         \big[\a'_0\ket{0}_4 + \a'_1\ket{1}_4 + \a'_2\ket{2}_4\big],  \\
%
&=&
\left[
\a_0\,\a'_0\,a^\dag_2 +
\a_0\,\dfrac{\a'_2}{\sqrt{2}}\, a^\dag_2 \,\left(a^\dag_4\right)^2+
\dfrac{\a_2}{\sqrt{2}}\,\a'_0\, \left(a^\dag_1\right)^2\,a^\dag_2 
\right. 
\left.
+ \; 
\dfrac{\a_2}{\sqrt{2}}\,\dfrac{\a'_2}{\sqrt{2}}\,\left(a^\dag_1\right)^2\,a^\dag_2 \,\left(a^\dag_4\right)^2
\right]\,
\ket{0,0,0,0}_{1234} \qquad  \label{Psi:0:w-CS:line1}\\
&+&
\left[
\a_0\,\a'_1\,a^\dag_2\,a^\dag_4 + 
\a_1\,\a'_0\,a^\dag_1\,a^\dag_2 +
\a_1\,\a'_1\,a^\dag_1\,a^\dag_2\,a^\dag_4 +
\a_1\,\dfrac{\a'_2}{\sqrt{2}}\, a^\dag_1\,a^\dag_2\,(a^\dag_4)^2
\right. \no
&{}&
\left.
\hspace*{2.75in}
+\, 
\dfrac{\a_2}{\sqrt{2}}\,\a'_1\, (a^\dag_1)^2\,a^\dag_2\,a^\dag_4
\right]\, \ket{0,0,0,0}_{1234}, \label{Psi:0:w-CS:line2} \\
&\equiv&
\ket{\Psi_{02';2'0}^{(0)}} + \ket{\Psi_{1,1'}^{(0)}}, \label{Psi:0:w-CS:line3}
\eea  
\esub
where the input state $\ket{\Psi^{(0)}} _{1234}$ has been separated into two branches.
\Eq{Psi:0:w-CS:line1} separates out that branch $\ket{\Psi_{02';2'0}^{(0)}}$ 
of the input state that contains only the 
 states $\ket{0}_k$ and  $\ket{2}_k$ in modes $k=1,4$.
\Eq{Psi:0:w-CS:line2} 
$\ket{\Psi_{1,1'}^{(0)}}$ separates out the remaining branch of the input state $\ket{\Psi^{(0)}} _{1234}$ 
that involve either input  states $\ket{1}_1$ , $\ket{1}_4$, or both.

In the following, we will first concentrate on transformation of the input state $\ket{\Psi_{02';2'0}^{(0)}}$ 
which after the measurement involves the single $3$-photon output state $\ket{1,1,0,1}_{1234}$, 
and only 5-photon accidental states.
Subsequently, we will analyze the transformation of the remaining input state $\ket{\Psi_{1,1'}^{(0)}}$, 
which after the measurement also involves the output state  $\ket{1,1,0,1}_{1234}$, 
but now with only 4-photon accidental states.

\subsection{Transformation of the input state $\ket{\Psi_{02';2'0}^{(0)}}_{1234}$}
After applying the $4\times 4$ unitary $\U = BS_{14}\cdot(NLPSG_{123}\otimes PHASE_4)$ on the mode operators,
as illustrated in \Fig{fig:direct:meas:nlpsg}, we have
(note: under $PHASE_4$ we have $\ket{n}_4 \to e^{i\,n\,\varphi}\,\ket{n}_4$ for $n\in\{0,1,2\}$).
\bsub
\bea{Psi:2:again}
\hspace{-0.75in}
\ket{ \Psi^{(0)}_{0,2';2,0'}}_{1234} & \xrightarrow{\trm{\tiny NLPSG}_{123}\otimes\trm{\tiny PHASE}_4}&   \ket{ \Psi^{(1)}_{0,2';2,0'} }_{1234}
\xrightarrow{BS_{14}}  \ket{ \Psi^{(2)}_{0,2';2,0'} }_{1234}, \no
&=&
\left[
\a_0\,\a'_0\, \sum_{j=1}^4\,a^\dag_j\,\U_{j2} +
\a_0\,\dfrac{\a'_2}{\sqrt{2}}\,e^{i\,2\,\varphi}\,  \left(\sum_{j=1}^4\,a^\dag_j\,\U_{j2}\right) \,\left(\cos(\theta/2)a^\dag_4+\sin(\theta/2)\,a^\dag_1\right)^2 
\right. \no
&+&
\dfrac{\a_2}{\sqrt{2}}\,\,\a'_0\,
\left(\sum_{j=1}^4\,a^\dag_j\,\U_{j1}\right)
\left(\sum_{k=1}^4\,a^\dag_k\,\U_{k1}\right)
\left(\sum_{\ell=1}^4\,a^\dag_\ell\,\U_{\ell 2}\right) \no
&+&
\left.
\frac{\a_2}{\sqrt{2}}\,\frac{\a'_2}{\sqrt{2}}\,e^{i\,2\,\varphi}\,
\left(\sum_{j=1}^4\,a^\dag_j\,\U_{j1}\right)
\left(\sum_{k=1}^4\,a^\dag_k\,\U_{k1}\right)
\left(\sum_{\ell=1}^4\,a^\dag_\ell\,\U_{\ell 2}\right)\,
\left(\cos(\theta/2)a^\dag_4+\sin(\theta/2)\,a^\dag_1\right)^2
\right],\qquad\; \label{T:states:opr:sum:cl-SPDC} \\
&\xrightarrow{\Pi^{(1234)}}&
 \ket{T_{0,2'}}_{1234} + \ket{T_{2,0'}}_{1234} + \ket{T_{2,2'}}_{1234}. \label{T:states:sum:cl-SPDC}
\eea
\esub
where in \Eq{T:states:opr:sum:cl-SPDC}  we have explicitly carried out the BS transformation on mode-$4$, 
$a^\dag_4 \rightarrow \cos(\theta/2)\,a^\dag_4+\sin(\theta/2)\,a^\dag_1$, (but \textit{not} on mode-$1$).
Additionally, we have explicitly implemented the the PHASE gate $I_{123}\otimes\,e^{i\,a^\dag_4\,a_4\,\varphi}$ on mode-$4$, 
which on states sends $\ket{0}_4 \rightarrow  \ket{0}_4$ and $\ket{2}_4 \rightarrow  e^{i\,2\,\varphi}\,\ket{2}_4$, 
and which we have incorporated by hand,
 having the net effect of sending $\a'_2\rightarrow \a'_2\,e^{i\,2\,\varphi}$.
Here the states $ \ket{T_{i,j'}}_{1234}$ listed in \Eq{T:states:sum:cl-SPDC}, 
arising from the transformation of the input state $\ket{i}_1\,\ket{j}_4$,
are those three or more photon states that survive under measurement projection. Recall that $p^{(k)}_0=0$ for for mode $k$, so that the states that remain after projection must contain three or more photons, with at least one photon in each of modes $k=\{1,2,4\}$.

The individual states are given by 
\bea{T:0:2prime:summary}
\ket{T_{0,2'}}_{1234} &=&
 \a_0\,\dfrac{\a'_2}{\sqrt{2}}\,e^{i\,2\,\varphi}\,\sin(\theta)\,\U_{22}\ket{1,1,0,1}_{1234}, \no
             &=& \a_0\,\dfrac{\a'_2}{\sqrt{2}}\,e^{i\,2\,\varphi}\,\sin(\theta)\,  S_{22}\ket{1,1,0,1}_{1234}, 
\eea
and
\bea{T:2:0prime:summary}
\hspace{-.6in}
\ket{T_{2,0'}}_{1234}
&=&\frac{\a_2}{\sqrt{2}}\,  \a'_0\,
\left[
\U_{11}\,\U_{21}\,\U_{42} + \U_{11}\,\U_{41}\,\U_{22} + \U_{21}\,\U_{11}\,\U_{42} + 
\U_{21}\,\U_{41}\,\U_{12} + \U_{41}\,\U_{11}\,\U_{22} + \U_{41}\,\U_{21}\,\U_{12} 
\right]\, \ket{1,1,0,1}_{1234}, \qquad\;\; \no
&=&
\frac{\a_2}{\sqrt{2}}\, \a'_0\,\sin(\theta) 
\left[
-S_{11}\,\left( S_{11}\,S_{22} + S_{21}\,S_{12} \right)
\right]\, \ket{1,1,01}_{1234},
\eea
where we recognize $\beta_0$ from \Eq{beta:0} as the amplitude in \Eq{T:0:2prime:summary}, and $\beta_2$ from \Eq{beta:2}
 as the  amplitude in \Eq{T:2:0prime:summary}.
Thus, combining the above two results 
and upon imposing Condition-0 \Eq{beta:0} and Condition-2  \Eq{beta:2} only, which ensures that $\beta_0=\beta_2\equiv\beta$,
we already have our \textit{primary} interference contribution, namely
\bea{T02prime:plus:T20prime:summary}
\ket{T_{0,2'}}_{1234} + \ket{T_{2,0'}}_{1234}
&=& 
\frac{1}{\sqrt{2}}\,\sin(\theta)\,
\left[
-S_{11}\,\left( S_{11}\,S_{22} + S_{21}\,S_{12} \right)\,\a'_0\,\a_2 + \a_0\,\a'_2\,e^{i\,2\,\varphi}\,S_{22}
\right] \, \ket{1,1,0,1}_{1234}, \\
&=&
\frac{1}{\sqrt{2}}\,\sin(\theta)\,\beta\,
\left[
\a'_0\,\a_2 + \a_0\,\a'_2\,e^{i\,2\,\varphi}
\right] \, \ket{1,1,0,1}_{1234}, \;\; \textrm{\textit{if} impose Conditions 0 \& 2}, \qquad\;\; \\
&\rightarrow &
\sqrt{2}\,\sin(\theta)\,\beta\,|\a_0|\,|\a_2|\,e^{i\,\varphi}\,\cos(\varphi)\,\ket{1,1,0,1}_{1234},  \quad \textrm{for}\; \a'_0=\a_0\; \textrm{\&}\;  \a'_2=\a_2.
\eea
This last term leads to a coincidence probability arising only from the $3$-photon state $\ket{1,1,0,1}_{1234}$ contribution:
\be{p:1234:3photon:level:summary}
P^{(3-\textrm{photons})}_{1234} \rightarrow 2\,\xi_1^2\,\xi_2^2\,\xi_4^2\,\sin^2(\theta)\,|\beta|^2\,|\a_0|^2\,|\a_2|^2\,\cos^2(\varphi),
\ee
where we have also included the finite detection efficiency factors.

Lastly, the remaining 5-photon states $\ket{T_{2,2'}}_{1234}$ generated from $\U$ acting on $\ket{ \Psi^{(0)}_{0,2';2,0'}}$, 
which \textit{do not} contribute to the above primary interference pattern are given by
\bea{T:2:2prime:summary}
\ket{T_{2,2'}}_{1234} &=&
\a_2\,\a'_2\,e^{i\,2\,\varphi} \, \Big( \no
&{}&
\ket{1,2,0,2}_{1234}\,
\left[
 \sin(\theta)\, \Ucs_{202}  +  \sin^2(\theta/2)\,\Uss_{202}
\right] \no
&+&
\ket{1,2,1,1}_{1234}\,
\left[
 \sin(\theta)\, \Ucs_{211} 
\right] \no
&+&
\ket{1,1,1,2}_{1234}\,
\left[
 \sin(\theta)\, \Ucs_{112}  +  \sin^2(\theta/2)\,\Uss_{112}
\right] \no
&+&
\ket{1,1,2,1}_{1234}\,
\left[
 \sin(\theta)\, \Ucs_{121} 
\right] \no
&+&
\ket{2,2,0,1}_{1234}\,
\left[
 \cos^2(\theta/2)\,\Ucc_{201} + \sin(\theta)\, \Ucs_{201} 
\right] \no
&+&
\ket{2,1,1,1}_{1234}\,
\left[
 \cos^2(\theta/2)\,\Ucc_{111} + \sin(\theta)\, \Ucs_{111} 
\right] \no
&+&
\left.
\ket{2,1,0,2}_{1234}\,
\left[
 \cos^2(\theta/2)\,\Ucc_{102} + \sin(\theta)\, \Ucs_{102}  +  \sin^2(\theta/2)\,\Uss_{102}
\right] \right),
\eea
where the various matrix elements $\{\Ucs_{202}, \Uss_{202}, \ldots\}$ in terms of $\U_{ij}$ 
are  listed explicitly in the Appendix \App{app:U:coeffs}.
(Note: $\{cc, cs, ss,\}$ superscripts indicate that terms are multiplied by 
$\{\cos^2(\theta/2),2\,\cos(\theta/2)\sin(\theta/2),\sin^2(\theta/2)\}$
and $\{c, s\}$ superscripts indicate terms are multiplied by $\{\cos(\theta/2),\,\sin(\theta/2)\}$.
The subscripts $i,j,k$ indicate that the amplitudes multiply the state $\ket{i,j,k}_{2,3,4}$).
The important point to note is that by containing  5-photon terms $\ket{T_{2,2'}}_{1234}$ is automatically orthogonal to the 3-photon state 
$\ket{1,1,0,1}_{1234}$ upon which the primary interference effects occurs. Additionally, each term in \Eq{T:2:2prime:summary} is multiplied by $e^{i\,2\,\varphi}$, and is also orthogonal to every other term in $\ket{T_{2,2'}}_{1234}$. Hence, upon squaring these amplitudes for the probability, these terms simply contribute to a (BS-angle dependent) dc accidental term $DC(\theta)$, 
independent of the phase angle $\varphi$.

\subsection{Transformation of the input state $\ket{\Psi_{1;1'}^{(0)}}_{1234}$}
Turning to the the transformation of the terms listed in \Eq{Psi:0:w-CS:line2} containing $\ket{1}_1$, $\ket{1}_4$ or both, we have
\bsub
\bea{Psi:2:w-CS:summary}
\hspace{-0.85in}
\ket{ \Psi^{(0)}_{1,1'}}_{1234} & \xrightarrow{\trm{\tiny NLPSG}_{123}\otimes\trm{\tiny PHASE}_4}&  \ket{ \Psi^{(1)}_{1,1'} }_{1234}
 \xrightarrow{BS_{14}}  \ket{ \Psi^{(2)}_{1,1'} }_{1234}, \no
&=&
\left[
\a_0\,\a'_1\,
\left(\sum_{j=1}^{4} a^\dag_j\,\U_{j 1}\right) \,
\left( \cos(\theta/2)\,a^\dag_4 +  \sin(\theta/2)\,a^\dag_1 \right) +
\a_1\,\a'_0\,
\left( \sum_{j=1}^{4} a^\dag_j\,\U_{j 1} \right)\, 
\left( \sum_{k=1}^{4} a^\dag_k\,\U_{k 2} \right)\, 
\right. \no
&{}&
\hspace{-.6in}
+\,
 \a_1\,\a'_1\,e^{i\,\varphi}\,
\left( \sum_{j=1}^{4} a^\dag_j\,\U_{j 1} \right) \, 
\left( \sum_{k=1}^{4} a^\dag_k\,\U_{k 2} \right)\,
\left( \cos(\theta/2)\,a^\dag_4 +  \sin(\theta/2)\,a^\dag_1 \right) \\
&{}& 
\hspace{-.6in}
+ \, 
\a_1\,\dfrac{\a'_2}{\sqrt{2}}\, e^{i\,2\,\varphi}\,
\left(\sum_{j=1}^4 a^\dag_j\U_{j 1}\right)\,
\left(\sum_{k=1}^4 a^\dag_k\U_{k 2}\right)\,
\left( \cos(\theta/2)\,a^\dag_4 +  \sin(\theta/2)\,a^\dag_1 \right)^2 \no
&{}&
\left.
\hspace{-.6in}
+\, 
\dfrac{\a_2}{\sqrt{2}}\,\a'_1\, e^{i\,\varphi}\,
\left(\sum_{j=1}^4 a^\dag_j\U_{j 1}\right)\,
\left(\sum_{k=1}^4 a^\dag_k\U_{k 1}\right)\,
\left(\sum_{\ell=1}^4 a^\dag_\ell\U_{\ell 2}\right)\,
\left( \cos(\theta/2)\,a^\dag_4 +  \sin(\theta/2)\,a^\dag_1 \right)
\right]\, \ket{0,0,0,0}_{1234},\qquad\;  \label{Psi:2:w-CS:line1}  \\
&{}& 
\hspace{-.6in}
\xrightarrow{\Pi^{(1234)}}
 \ket{T_{1,1'}}_{1234} + \ket{T_{1,2'}}_{1234}  + \ket{T_{2,1'}}_{1234}. 
\eea
\esub
Following the same procedure as above, the state that survives after measurement projection
and contributes to the primary coincidence interference effect is
\bea{T:1:1prime:}
\ket{T_{1,1'}}_{1234}  &=& 
 \a_1\,\a'_1\,e^{i\,\varphi}\,
\ket{1,1,0,1}_{1234}\,
\left[
\cos(\theta/2)\,\Ucprime_{101} +  \sin(\theta/2)\,\Usprime_{101}\,
\right], \no
&=& \a_1\,\a'_1\,e^{i\,\varphi}\, \ket{1,1,0,1}_{1234}\; \beta_1 \cos(\theta),
\eea
which arises from the transformation of the input state $\ket{1}_1\,\ket{1}_4$.
In the above we have defined 
\be{beta1}
\beta_1  \equiv S_{11}\,S_{22} + S_{21}\,S_{12} \myover{\longrightarrow}{\beta\to\beta_{max}=1/2} \, \beta_{max}=1/2, \quad (\beta_{max}^2=1/4),
\ee
where
$\beta_1\to \beta=1/2$ 
\textit{if} we were to  impose Condition-1, \Eq{beta:1} in addition to the previously imposed 
Condition-0,  \Eq{beta:0} and Condition-2,  \Eq{beta:2}, 
which would then make $\beta^2\to\beta^2_{max}=1/4$.

The remaining 4-photon orthogonal accidental states arising from the transformation of the 
input states $\ket{1}_1\,\ket{2}_4$ and  $\ket{2}_1\,\ket{1}_4$ are given by
\bea{T1:2prime:2:1prime:summary}
\hspace{-1in}
\ket{T_{1,2'}}_{1234}  &+& \ket{T_{2,1'}}_{1234} = \no
&{}&
\hspace{-.8in}
\Big(
e^{i\,\varphi}\,\ket{1,2,0,1}_{1234}\,
\left[
\a_1\,\a'_2\,e^{i\,\varphi}\,
\sin(\theta)\, \Ucsprime_{201}  
+
\a_2\,\a'_1\,
\left( \cos(\theta/2)\, \Ucprime_{201}  +  \sin(\theta/2)\,\Usprime_{201}\right)
\right] 
\no
&{}&
\hspace{-.8in}
+\,
e^{i\,\varphi}\,\ket{1,1,1,1}_{1234}\,
\left[
\a_1\,\a'_2\,e^{i\,\varphi}\,
\sin(\theta)\, \Ucsprime_{111}  
+
\a_2\,\a'_1\,
\left( \cos(\theta/2)\, \Ucprime_{111}  +  \sin(\theta/2)\,\Usprime_{111}\right)
\right]
\no
&{}&
\hspace{-.8in}
+\,
e^{i\,\varphi}\,\ket{1,1,0,2}_{1234}\,
\left[
\a_1\,\a'_2\,e^{i\,\varphi}\,
\left( \sin(\theta)\, \Ucsprime_{102}  +  \cos^2(\theta/2)\,\Uccprime_{102}\right)
+
\a_2\,\a'_1\,
\left( \cos(\theta/2)\, \Ucprime_{102}  +  \sin(\theta/2)\,\Usprime_{102}\right)
\right] 
\no
&{}&
\hspace{-.8in}
+\,
e^{i\,\varphi}\,\ket{2,1,0,1}_{1234}\,
\left[
\a_1\,\a'_2\,e^{i\,\varphi}\,
\left( \sin(\theta)\, \Ucsprime_{101}  +  \sin^2(\theta/2)\,\Ussprime_{101}\right)
+
\a_2\,\a'_1\,
\left( \cos(\theta/2)\, \Ucprime_{101}  +  \sin(\theta/2)\,\Usprime_{101}\right)
\right] \Big), \qquad\;\;
\eea
Again, the various matrix elements $\{\U'^{(cs)}_{202}, \U'^{(ss)}_{202}, \ldots\}$ in terms of $\U_{ij}$ 
are  listed explicitly in \App{app:U:coeffs}.
Note that squaring each of the above amplitudes will generate a $\cos(\varphi)$ higher-order interferences in the accidentals.

\subsection{Form of the unnormalized post measurement state}
The complete output state upon transformation by $\U$  
for a general w-CS input states on mode 1 and 4 is then
\bea{full:state:after:U:for:csSPDC:w-CS:input:summary}
\ket{\Psi^{(0)}}_{1234} \xrightarrow{\U} \ket{\Psi^{(2)}}_{1234}  &\equiv& \ket{T_{0,2'}}_{1234} + \ket{T_{2,0'}}_{1234} + \ket{T_{2,2'}}_{1234}  ,\no
&+&  \ket{T_{1,1'}}_{1234} + \ket{T_{1,2'}}_{1234}  + \ket{T_{2,1'}}_{1234}.
\eea
where the top line comes from the transformation of 
$\ket{\Psi_{02';20}^{(0)}}$ and 
the bottom line arises from the transformation of  
$\ket{\Psi_{1,1'}^{(0)}}$.
Recall that the state after projection is given by
\bea{tilde:Psi:3}
\ket{\tilde{\Psi}^{(2)}}_{1234} &\equiv& \Pi^{1234}\,\ket{\Psi^{(2)}}_{1234} 
= \hspace{-.2in} \sum_{n,m,r,s= 0}^{4\;\prime}\,  p^{(1)}_{n}\,p^{(2)}_{m}\,\left(1-p^{(3)}_{r}\right)\,p^{(4)}_{s}\,\ket{n,m,r,s}_{1234}
\IP{n,m,r,s}{\Psi^{(2)}}_{1234}, \qquad 
\eea
where the prime on the summation indicates that we are in the approximation that each mode contains at most two photons.
Since $p^{(k)}_0 = 0$, only states with at least one photon in modes $k\in\{1,2,4\}$ survive the measurement projection, and therefore 
$\ket{\tilde{\Psi}^{(2)}}_{1234}$ contains the 3-photon state  $\ket{1,1,0,1}_{1234}$, plus 4- and 5-photon accidental states that also contribute to the coincidence counts when detectors with finite detection efficiencies are employed.

The primary coincidence interference term arises from the $\ket{1,1,0,1}_{1234}$ portion of $\ket{\tilde{\Psi}^{(2)}}_{1234}$ 
which has the form
\be{Psi:3:form:summary}
\hspace{-.6in}
\ket{\tilde{\Psi}^{(2)}} =  
(\textrm{prefactor})^{1/2}\times\,
\left[
\beta\,f_3(\theta,\varphi)\,\ket{1,1,0,1}_{1234} +
\alpha\, \ket{\tilde{\Psi}_4^{(2)}(\theta,\varphi)}_{1234}  +
\alpha^2 \, \ket{\tilde{\Psi}_5^{(2)}(\theta)}_{1234} 
\right], 
\ee
where we have defined the prefactor as 
\be{prefactor:defn}
(\textrm{prefactor})^{1/2} = \dfrac{\xi_1\,\xi_2\,\xi_4\,\alpha^2\,e^{i\varphi}}{1+\alpha^2+\alpha^4/2} 
 = \dfrac{\xi_1\,\xi_2\,\xi_4\,\bar{n}\,e^{i\varphi}}{1+\bar{n}+\bar{n}^2/2}, 
\ee
where $\bar{n}\approx \alpha^2$ is the mean number of photons in the w-CS. 
Additionally, we define the interference amplitude $f_3(\theta,\varphi)$, after factoring out $\beta$, as
\bea{f3:defn}
\hspace{-.5in}
\textrm{\textit{Interference Amplitude}:} &{}& \no
&{}&
\hspace{-1.5in}
\trm{(i)}\; 
f_3(\theta,\varphi) = \sin(\theta)\,\cos(\varphi) +  (\beta_1/\beta)\,\cos(\theta), 
\textrm{ \textit{only assuming} Condition-$0$ \& Condition-$2$, i.e\;\;} \beta_0=\beta_2\equiv\beta, \qquad \label{f3:defn:1}\\
&{}&
\hspace{-1.5in}
\trm{(ii)}\; 
f_3(\theta,\varphi)\to \sin(\theta)\,\cos(\varphi) +  \cos(\theta),\; \hspace{.4in} \tit{additionally} \textrm{ imposing Condition-$1$, i.e.\;} 
 \beta_0=\beta_1=\beta_2\equiv\beta. \label{f3:defn:2}
\eea
Here $\ket{\tilde{\Psi}_4^{(2)}(\theta,\varphi)}_{1234} $ and $\ket{\tilde{\Psi}_5^{(2)}(\theta)}_{1234} $ are (unnormalized state) contributions from the $4$-photon and $5$-photon states respectively, that contribute to the accidentals, and we have used $\a'_k=\a_k =\dfrac{\alpha^k}{\sqrt{1+\alpha^2+\alpha^4/2}}$ in \Eq{prefactor:defn} for simplicity. 
$\theta$ is the BS angle (with $\theta=\pi/2$ for a 50:50 BS), and $\varphi$ is the phase shift angle in mode-$4$.
The final interference probability $P_{1234}$, imposing all three Conditions-0,1,2 then has the form 
\bsub
\bea{P:1234:All:summary}
P_{1234} &=&  ||\,\ket{\tilde{\Psi}^{(2)}}_{1234}||^2 = 
\left[\dfrac{ \xi^2_1\, \xi^2_2\, \xi^2_4\, \bar{n}^2 }{ (1+\bar{n}+\bar{n}^2/2)^2} \right]\,
\Big[
\beta^2\,f^2_3(\theta,\varphi) + \bar{n}\,AC(\theta,\varphi) + \bar{n}^2\,DC(\theta)
\Big], \label{P:1234:All:line1} \\
&{}&
\bar{n} = \alpha^2, \quad
AC(\theta,\varphi) = {}_{1234} \IP{\tilde{\Psi}_4^{(2)}}{\tilde{\Psi}_4^{(2)}}_{1234}   \quad
DC(\theta) =  {}_{1234} \IP{\tilde{\Psi}_5^{(2)}}{\tilde{\Psi}_5^{(2)}}_{1234}. \label{P:1234:All:line2}
\eea
\esub
Note that the first and third terms in the right square brackets of \Eq{P:1234:All:line1}  arise from input states on mode-$1$ and $4$ that contain only 0 and 2 photons when a 50:50 BS ($\theta=\pi/2$) is used (i.e. $\cos(\theta)\to0$ wipes out the interference contributions arising from the addition of the $\ket{1}_1$ and $\ket{1}_4$ input states). 
This is of course, just the well known HOM BS-induced interference effect in the context of our NLPSG MZI \cite{HOM:1987,Hach:2014}.
Also note that the first (interference) term in \Eq{P:1234:All:line1} is of
$\Order{\beta^2}\sim\Order{1}$, while each additional (accidentals) term scales as 
$\Order{\bar{n}}\ll1$ and $\Order{\bar{n}^2}\lll1$, respectively. 
\Eq{P:1234:All:line1} with \Eq{P:1234:All:line2} is one of the main results of this work, to which we will now specialize to both the KLM and MRR implementation of the NLPSG.

\begin{figure}[ht]
\begin{tabular}{cc}
\hspace{-.35in}
\includegraphics[width=3.75in,height=2.15in]{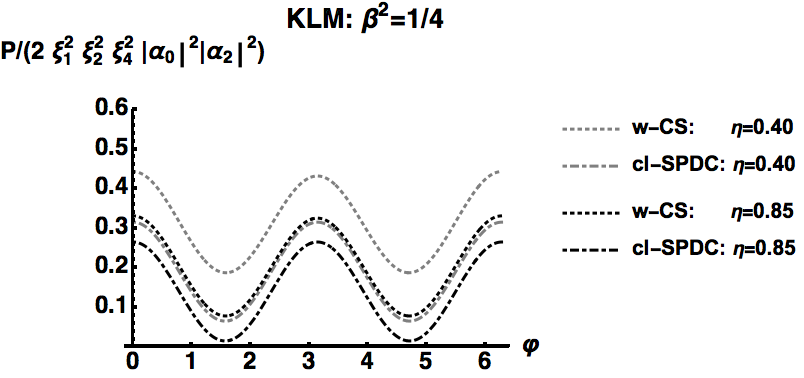} &
\includegraphics[width=3.75in,height=2.15in]{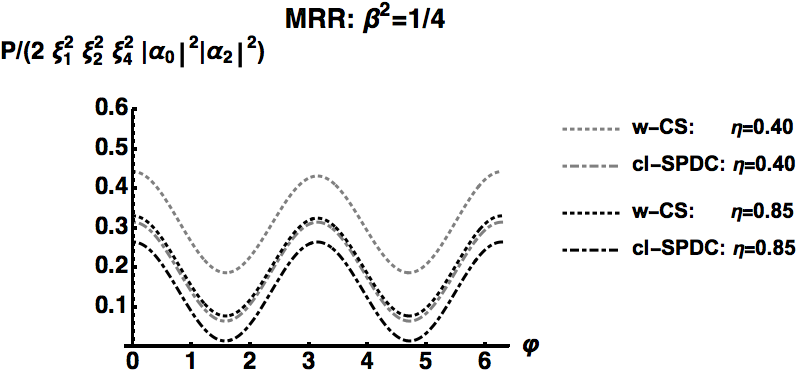}
\end{tabular}
\caption{Plot of $P'_{1234}$ from \Eq{P:1234:All:summary}
for the 
(left) KLM,
(right) MRR
NLPSG
for 
$\beta^2=1/4$
for (dotted) w-CS input states and (dot-dashed) cl-SPDC input states 
with  detection efficiencies  (gray) $40\%$ and  (black) $85\%$.
The left and right graphs are \textit{identical}.
}\label{fig:Pprime1234:KLM:MRR:betaSqrd:quarter:det:eff:40:85:summary}
\end{figure}

In \Fig{fig:Pprime1234:KLM:MRR:betaSqrd:quarter:det:eff:40:85:summary} we plot
the scaled probability  $P^{\prime(\theta=\pi/2,\varphi)}_{1234}$ 
(i.e. defined from \Eq{P:1234:All:summary} as  $P_{1234} ~\equiv ~\textrm{prefactor}\times P'_{1234}$ ) 
for  coincidences 
(left) KLM,
(right) MRR
using  co-linear SPDC (cl-SPDC, dashed) and weak coherent (w-CS, solid) input states  
with a 50:50 BS ($\theta=\pi/2$), and $|\a_2/\a_0|^2=0.1$
with finite detection efficiencies (gray, black) $\xi_1=\xi_2=\xi_4 \equiv \xi = \{0.40,0.85\}$,
at the optimal reflection coefficients
$r^{*2}_1=r^{*2}_3,\, r^{*2}_2 \Rightarrow |\beta|^2 = 1/4$.
The (left) KLM and (right) MRR curves are identical.
The reason these curves are \textit{identical}, is that even though  $S^{(MRR)}$ and $S^{(KLM)}$
are not strictly identical, i.e.  $S^{(MRR)} \ne S^{(KLM)}$, 
they are \textit{effectively identical} in the sense that the upper left $2\times 2$ sub-matrix
$
\small{
\left(\begin{array}{cc} S_{11} & S_{21} \\S_{21} & S_{22}\end{array}\right)
}
$,
of each unitary matrix
are identical at $\beta^2=1/4$, which now enforces Condition-1, along with Condition-0 and Condition-2 which were previously satisfied, 
while the third row and third column of the each unitary matrix are different.
This is how the MRR-NLPSG encompasses the KLM-NLPSG (since the former solution was modeled after the latter's).
This is \textit{not} the case at other values of  $\beta^2\ne 1/4$. 

\begin{figure}[ht]
\begin{tabular}{c}
\hspace{1in}
\includegraphics[width=4.0in,height=2.5in]{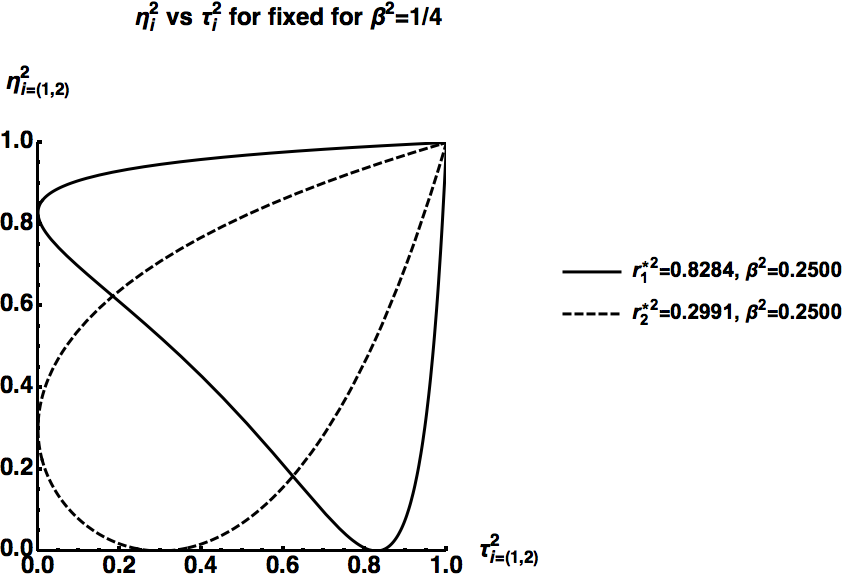} 
\end{tabular}
\caption{
Plots of the physical MRR  transmissivities  
(solid) $\eta^2_{1=3}$ vs $\tau^2_{1=3}$, and
(dashed) $\eta^2_{2}$ vs $\tau^2_{2}$
for \textit{fixed} values of the 
fictitious KLM reflectivities $r^{*2}_1,\, r^{*2}_2$ yielding $\beta^2=1/4$.  
}\label{fig:etaSqrd:tauSqrd:fixed:t2Sqrd:betaSqrd:in:body:of:manuscript}
\end{figure}

The new feature using the MRR-NLPSG is the one-dimensional manifold relationship 
between the \textit{phyisical transmission} coefficients 
$\tau_i$ and $\eta_i  = \eta_i(\tau_i)$ of the MRR NLPSG in terms of  fictitious KLM  \tit{ effective refection coefficients} $r_i$
as described in \App{app:KLM:MRR:NLPSG}, and discussed more fully in \cite{Alsing_Hach:2019}. 
That is, by modeling the solutions of the MRR NLPSG \tit{as if} it were composed of three KLM BS, 
one finds the MRR solutions for the \tit{fictitious KLM} $r^*_i$ that yield $\beta^2=1/4$ 
define a 1-parameter family (manifold) of \tit{physical MRR transmission} coefficients $\eta_i = \eta_i(\tau_i; r^*_i)$ 
(this is true \textit{in general} regardless of the value of $r_i\Rightarrow\beta^2$ considered) 
given by
 \be{eta:tau:t}
 \eta_i(\tau_i; r^*_i)= \dfrac{r^*_i+ \tau_i}{1+r^*_i\, \tau_i}, \quad  |\t^*_i| \le 1 \Rightarrow  |\eta^*_i| \le 1,
  \quad \textrm{for fixed} \quad  |r^*_i| \le 1. 
 \ee 
 as shown in \Fig{fig:etaSqrd:tauSqrd:fixed:t2Sqrd:betaSqrd:in:body:of:manuscript}.
This affords a much greater freedom in the use of the physical transmission coefficients to realize the coincidence interference effect, 
over the single \textit{point-solution} obtained from the KLM-NLPSG. 

All the graphs for a 50:50 BS ($\theta=\pi/2$) have the same qualitative form 
$P'_{1234} = a_0 + a_1\,\cos(\varphi) + \beta^2\cos^2(\varphi)$ (see \Tbl{tbl:P1234:coeffs}).
We define the coincidence probability as
$P_{1234} ~= ~\textrm{prefactor}\times P'_{1234}$.
Here, the prefactor (see \Tbl{tbl:prefactor:visibilities}) scales as $\xi^6\,\bar{n}^2$ where 
$\bar{n}=\alpha^2 = \a^2_1=\sqrt{2}\,\a_2$ is the mean number of photons in the weak coherent state (w-CS).
The upward displacement of the probability curve indicates a larger value of the accidentals - essentially a DC noise offset.
In general, the higher the detection efficiency, the lower the noise floor, and the closer the curve nearly 
touches the abscissa, and consequently, the higher the visibility, as shown in \Tbl{tbl:prefactor:visibilities}.
%
\begin{table}
\begin{tabular}{|c|c|c|}\hline
\multicolumn{3}{|c|}{\bf{Coincidence Detection Probability}} \\ \hline
\multicolumn{3}{|c|}{$P'_{1234} = a_0 + a_1\,\cos(\varphi) + \beta^2\,\cos^2(\varphi)$} \\ \hline \hline
\multicolumn{3}{|c|}{$(a_0, a_1)\;\textrm{for}\;\beta^2 = 1/4$} \\ \hline
\; $\xi\,$\textbackslash \it{input state}\;  & \;\it{cl-SPDC}\; & \;\it{w-CS}\; \\ \hline
0.40 &  (0.065, 0.000) & (0.188, 0.000) \\ \hline
0.85 & (0.015, 0.006) & (0.078, 0.003) \\ \hline
\end{tabular}
\caption{Form of the scaled coincidence interference probability $P'_{1234} = a_0 + a_1\,\cos(\varphi) + \beta^2\,\cos^2(\varphi)$
where the full probability \Eq{P:1234:All:summary} is given by  $P_{1234} = \textrm{prefactor}\times P'_{1234}$ (see \Tbl{tbl:prefactor:visibilities}). }
\label{tbl:P1234:coeffs}
\end{table}
%
\begin{table}
\begin{tabular}{|c|c|c|}\hline
\multicolumn{3}{|c|}{\bf prefactor} \\ \hline \hline
\multicolumn{3}{|c|}{$\;P_{1234} = \textrm{prefactor}\times P'_{1234}\;$} \\ \hline
\; $\xi\,$\textbackslash\,\it{input state}\;  & \;\it{cl-SPDC}\; & \;\it{w-CS}\; \\ \hline
0.40 &  \;$6.7\times 10^{-4}$\; & \;$4.1\times 10^{-4}$\; \\ \hline
0.85 &  \;$6.2\times 10^{-2}$\; & \;$3.8\times 10^{-2}$\; \\ \hline
\end{tabular}
%
\hspace{0.5in}
\begin{tabular}{|c|c|c|}\hline
\multicolumn{3}{|c|}{\bf Visibilities} \\ \hline \hline
\multicolumn{3}{|c|}{$\beta^2 = 1/4$} \\ \hline
\; $\xi\,$\textbackslash \it{input state}\;  & \;\it{cl-SPDC}\; & \;\it{w-CS}\;\; \\ \hline
0.40 &  65\% & 41\% \\\hline
0.85 & 89\% & 65\% \\ \hline
\end{tabular}
\caption{
(left) prefactor (overall strength of the coincidence interference probability: $P_{1234} = \textrm{prefactor}\times P'_{1234}$,
(right) Visibilities of coincidence interference probability $P'_{1234}$ for input states cl-SPDC and w-CS for  $\beta^2 = 1/4$.}
\label{tbl:prefactor:visibilities}
\end{table}
In both the (left) and (right) figures of \Fig{fig:Pprime1234:KLM:MRR:betaSqrd:quarter:det:eff:40:85:summary} we note that using the 
cl-SPDC input states at the \textit{lower} detection efficiency of $40\%$  produces nearly the identical curve as using 
w-CS input states at the much \textit{higher} detection efficiency of $85\%$.

Note, if we generate input states at a rate $r_{states}$ states/sec and integrate for a time $T$, then the number of counts is given by $N_{counts} = \textrm{prefactor}\times r_{states}\times T$ for each of the $N_{\varphi}$ discrete values of $\varphi$ sampled (at minimum $10$). This implies that the total time to conduct the experiment will be on the order of $T_{exp}~\sim~N_{\varphi}\,N_{counts}/(\textrm{prefactor}\times r_{states})$, highlighting the implication of the higher detection efficiency increasing the value of the prefactor, 
thus reducing $T_{exp}$. Note that prefactor scales as $\xi_1^2\, \xi_2^2\,\xi_4^2\sim\xi^6$ so that a change in detection efficiency from $40\%$ to $85\%$ yields an increase of $(0.85/0.40)^6=92\sim 100$X
 in the strength of the effect, while also reducing the strength of the accidentals by $\sim 4$X.
 The use of more efficient detectors is clearly evident in \Tbl{tbl:prefactor:visibilities}.

\section{Conclusion}\label{sec:Conclusion}
In this work we have presented a direct MZI  interferometric coincidence test of the KLM and MRR NLPSG 
for detectors with finite detection efficiencies. 
In the past, the KLM NLPSG was tested indirectly through the use of two of them to form the basis of a CNOT gate. Essentially, this was a HOM interference on the two-photon branch of the input state (mode 1). Here we propose a straightforward HOM interference setup with a w-CS input state in each arm of a MZI, one arm containing the KLM or MRR NLPSG and the other arm containing a phase shifter. For a 50:50 BS, we show that the primary coincidence interference effect that appears on the ideal NLPSG ``success state" arises from the vacuum and two-photon mixing on the final MZI BS, a manifestation of the HOM effect. 
To make this calculation more experimentally relevant, we keep all terms in the MZI unitary transformation containing up to two photons in each of the four possible modes (three for the NLPSG in one arm of the MZI and one for the phase shifter in the other arm), so that we can include the accidentals that contribute to the coincidence measurement when detectors with finite efficiencies are employed. We further show how the MRR NLPSG encompasses the KLM NLPSG and utilizes the latter's maximum success probability fixed point solution as a parameter in a one dimensional manifold relationship between the physical transmissivities of the each MRR (that now replaces each KLM BS). 
Lastly, we additionally show that if one instead uses cl-SPDC input states in each arm of the MZI, where the single photon branch is absent, then one obtains qualitatively the same coincidence interference probability, however now with accidentals down by the square of the mean number of photons in the input state, and with a moderately increased interference visibility. While the generation of w-CS is much less resource intensive than that for the production of cl-SPDC states (with corresponding a higher generation rate), both types of inputs states can be utilized to validate the sign-flip by the measurement-induced NLPSG. Both of these approaches could be utilized in current photonic integrated waveguide devices, and experimental verification of these approaches are the focus of follow-on research.

\clearpage
\newpage
\appendix
\section{The KLM and MRR implementation of the NLPSG}\label{app:KLM:MRR:NLPSG}
\subsection{The KLM NLPSG}
\begin{figure}[ht]
\includegraphics[width=5.0in,height=2.0in]{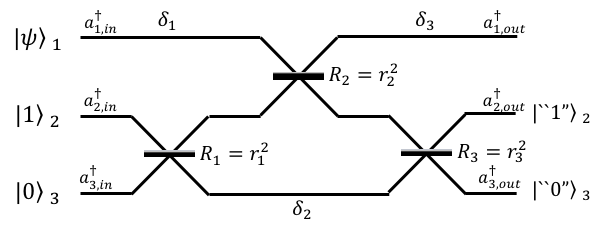}
\caption{The KLM NLSG using three ordinary beam splitters of reflectivities $R_i = r_i^2$, and optical path delays of $\delta_i$.
Mode 1 is the primary input state in a weak coherent state (w-CS) containing up to two photons.
Mode 2 and 3 are ancilla modes, initially in the state $\ket{1,0}_{23}$.
Success of NLPSG is heralded by the detection of the output ancilla modes in their initial state.
}\label{fig:skaar:nlsg}
\end{figure}

The KLM implementation of the NLPSG as shown in \Fig{fig:skaar:nlsg} utilized three individual BS of the form
\be{M:skaar}
\hspace{-.3in}
M_1 =
\left(
  \begin{array}{cc}
    r_1\,e^{i\phi_1} & \sqrt{1-r^2_1} \\
    \sqrt{1-r^2_1} & -r_1 \,e^{-i\phi_1} \\
  \end{array}
\right), \;
M_2 =
\left(
  \begin{array}{cc}
    -r_2\,e^{-i\phi_2} & \sqrt{1-r^2_2} \\
    \sqrt{1-r^2_2} & r_2 \,e^{i\phi_2} \\
  \end{array}
\right),\;
M_3 =
\left(
  \begin{array}{cc}
    r_3\,e^{i\phi_3} & \sqrt{1-r^2_3} \\
    \sqrt{1-r^2_3} & -r_3 \,e^{-i\phi_3} \\
  \end{array}
\right), 
\ee
with  real  BS \tit{reflection} coefficients  $-1\le r_i \le 1$ 
(\tit{reflectivities} $R_i = r^2_i$) 
\footnote{In \cite{Alsing_Hach:2019,  Alsing_Hach:2018}  we labeled the $r_i$ in this work as $t_i$, and called the later \tit{transmission coefficients}. We followed the calculation of Skaar \cite{Skaar:2004} so that the $t_i$ were in fact actually \tit{reflection} coefficients. All the calculations and conclusions in \cite{Alsing_Hach:2019,Alsing_Hach:2018} are uneffected, since the KLM $t_i$ functioned merely as parameters that defined the 1-dimensional manifold relationship between the \tit{physical transmission} coefficients $\eta_i$ and $\tau_i$  of MRR$_i$, see \Eq{eta:tau:t} and \Eq{eta:tau:t:summary:app}.}.
Note that we have chosen a (non-standard) matrix representation of the $2\times2$ BS matrix that contains only real coefficients \cite{Skaar:2004} such that $\det(M_i) = -1$ for $i\in\{1,2,3\}$.

Recall that a unitary transformation $\U=B_3\,B_2\,B_1$  affects the following transformations on the boson creation operators \cite{Skaar:2004,Alsing_Hach:2019,Alsing_Hach:2018}
\bea{B1B2B3:to:S}
a^\dag_i &\xrightarrow{B_1}& \sum_{j=1}^4\, a^\dag_j\,\,(B_{1})_{ji}
               \xrightarrow{B_2} \sum_{j=1}^4\,  \sum_{k=1}^4\, a^\dag_k \,(B_2)_{kj} \,(B_{2})_{ji}
               \xrightarrow{B_3} \sum_{j=1}^4\,  \sum_{\ell=1}^4\, a^\dag_\ell\,(B_{3})_{\ell k} \,(B_2)_{kj} \,(B_{1})_{ji} \equiv  \sum_{k=1}^4\, a^\dag_\ell\,\U_{\ell i}, \no
&{}&    
\hspace{-.6in} 
   \textrm{with}\quad  \U_{\ell i} = \sum_{k=1}^4\, \sum_{j=1}^4\,(B_{3})_{\ell k} \,(B_2)_{kj} \,(B_{1})_{ji} 
   \equiv (B_3\,B_2\,B_1)_{\ell i},
\eea
with $B_1$ acting first,  $B_2$ acting second, and  $B_3$ acting third as we traverse the NLPSG in \Fig{fig:skaar:nlsg} from left to right.

Putting the above three blocks together, we have the full evolution from left to right in \Fig{fig:skaar:nlsg}
\bsub
\bea{full:left:to:right:eqns:Skaar}
\left(
  \begin{array}{c}
    a^\dag_{1, out} \\
    a^\dag_{2, out}  \\
    a^\dag_{3, out} \\
  \end{array}
\right)^T &=&
\left(
  \begin{array}{c}
    a^\dag_{1, in}  \\
    a^\dag_{2, in} \\
    a^\dag_{3, in}  \\
  \end{array}
\right)^T\,
 (B_3\,B_2\,B_1)\, \\
&=&
\left(
  \begin{array}{c}
    a^\dag_{1, in}  \\
    a^\dag_{2, in} \\
    a^\dag_{3, in}  \\
  \end{array}
\right)^T \, 
\left(
\begin{array}{c|}   e^{i\,\delta_3} \\   {\begin{array}{cc}\hline 0 \\[-0.5em] 0 \end{array}} \end{array}
\begin{array}{cc}  0 &  \hspace*{-10pt} 0 \\ \hline {\begin{array}{c}   \hspace*{10pt}   \raisebox{-5pt}{${M}_3$} \vspace*{15pt}  \end{array}} \end{array}
\right)
%
\left(
\begin{array}{c|}   \\[-1.25em] {M}_2 \\[0.80em]  {\begin{array}{cc}\hline 0 & 0 \end{array}} \end{array}
\begin{array}{c} 0 \\[-.35em] 0 \\ {\begin{array}{c}\hline e^{i\,\delta_2} \end{array}} \end{array}
\right)
%
\left(
\begin{array}{c|}   e^{i\,\delta_1} \\   {\begin{array}{cc}\hline 0 \\[-0.5em] 0 \end{array}} \end{array}
\begin{array}{cc}  0 &  \hspace*{-10pt} 0 \\ \hline {\begin{array}{c}   \hspace*{10pt}   \raisebox{-5pt}{${M}_1$} \vspace*{15pt}  \end{array}} \end{array}
\right)
%
\\
&\equiv&
\left(
  \begin{array}{c}
    a^\dag_{1, in}  \\
    a^\dag_{2, in} \\
    a^\dag_{2, in}  \\
  \end{array}
\right)^T \, S \\
&=&
\left(
  \begin{array}{c}
    a^\dag_{1, in}  \\
    a^\dag_{2, in} \\
    a^\dag_{3, in}  \\
  \end{array}
\right)^T\,
\left(
  \begin{array}{ccc}
    S_{11} & S_{12} & S_{13} \\
    S_{21} & S_{22} & S_{23} \\
    S_{31} & S_{32} & S_{33} \\
  \end{array}
\right),\,
\label{aL_eq_S_aR}
\eea
\esub
where the superscript $T$ indicates the transpose (i.e. the matrix $\U = B_3\,B_2\,B_1$ acts on the row vector 
$(a^\dag_{1, in},  a^\dag_{2, in},  a^\dag_{3, in})$ from the right, as in \Eq{B1B2B3:to:S}).
The above product of BS defines the matrix $S$ representing the three mode (1,2,3) KLM NLPSG with components $S_{ij}$ 
(obtained by explicitly multiplying out $B_3\,B_2\,B_1$)
routing a photon initially in mode $j$ (second index) into the mode $i$ (first index). 
Here, the $\delta_i$ represent phase shifts due to the optical path length delays to and from the BSs.

Without loss of generality,  we will  henceforth only consider the simple case when all phases $\{\phi_i, \delta_i\}$ are identically zero.
This yields the $S$-matrix
\be{SKLM}
S^{(KLM)} = 
\left(\begin{array}{ccc}
 -r_2 
 & \dfrac{\sqrt{1-r_2^4} }{\sqrt{1+ r_2 + r_2^2 - 3\,r_2^3}}\      
 & \dfrac{\sqrt{1-r_2^2}\,\sqrt{r_2-3\,r_2^3}}{\sqrt{1+ r_2 + r_2^2 - 3\,r_2^3}} \\
  & & \\
    \dfrac{\sqrt{1-r_2^4} }{\sqrt{1+ r_2 + r_2^2 - 3\,r_2^3}} 
&  \dfrac{2\,r_2\,(1+r_2)}{1+ 2\, r_2 +  3\,r_2^3}  
& -\dfrac{\sqrt{1+r_2^2}\,\sqrt{r_2-3\,r_2^3}}{1+ 2\, r_2 +  3\,r_2^3}  \\
  & & \\
    \dfrac{\sqrt{1-r_2^2}\,\sqrt{r_2-3\,r_2^3}}{\sqrt{1+ r_2 + r_2^2 - 3\,r_2^3}}  
& -\dfrac{\sqrt{1+r_2^2}\,\sqrt{r_2-3\,r_2^3}}{1+ 2\, r_2 +  3\,r_2^3} 
&. \dfrac{1+ r_2 + r_2^2 + 3\,r_2^3}{1+ 2\, r_2 +  3\,r_2^3} \\
\end{array}\right)
\ee
Here we have imposed only Condition-0 and Condition-2 so that $\beta_0=\beta_2\equiv\beta$ so that 
\bsub
\bea{beta:KLM}
r_1(r_2)=r_3(r_2) &=&  \dfrac{\sqrt{1+r_2^2}}{\sqrt{(1-r_2)\,(1+2\,r_2 + 3\,r_2^2)}},  \label{r1:of:r2} \\
\beta(r_2) &=& \dfrac{2\,r_2\,(1+r_2)}{1+ 2\, r_2 +  3\,r_2^3} = S_{22}. \label{beta:of:r2}
\eea
Maximizing $\beta$ in  \Eq{beta:of:r2} over $r_2$ yields the optimal operating values \cite{Skaar:2004,Alsing_Hach:2019,Alsing_Hach:2018}
\bea{r1:r2:beta:star:KLM}
                                 r_2^*  &=& \sqrt{2} - 1 \hspace{.35in}= 0.424214,         \qquad r_2^{* 2} = 0.171573, \\
\trm{KLM:}\qquad  r^*_1=r^*_3 &=& \dfrac{1}{\sqrt{4 - 2\,\sqrt{2}}}\;\; = 0.92388, \qquad \;r_1^{* 2} =  r_3^{* 2} = 0.853553, \\
                                 \beta  &=& \dfrac{1}{2},                                                      \qquad \hspace{1.45in}|\beta|^2 = \dfrac{1}{4}, 
\eea
\esub
%
with maximum NLPSG success probability $|\beta|^2_{max} = 1/4$.
Note that due to terms the linear in $r_2$ in \Eq{r1:of:r2} and \Eq{beta:of:r2} and  $r^2_1(-r_2)\ne r_1^2(r_2)$, we have
similarly $|\beta(-r_2)|^2\ne |\beta(-r_2)|^2$. For example, while $|\beta(-r^*_2)|^2=1/2$, 
we have $r^2_1(-r^*_2) >1$, and hence this unphysical solution must be rejected.

\subsection{The MRR NLPSG}
We now wish to extend the above considerations for the KLM version of the  NLPSG to the MRR version \cite{Alsing_Hach:2019,Alsing_Hach:2018} 
by replacing each KLM BS by a MRR. 
Each MRR$_i$  now has an upper and lower \textit{transmission} coefficient $\eta_i,\tau_i$, 
 phase angle $\theta_i$, and waveguide bus delays $\delta_i$ for $i\in\{1,2,3\}$. 
 In \cite{Alsing_Hach:2019,Alsing_Hach:2018}  the authors modeled the solutions of the MRR NLPSG by treating each MRR element 
 \textit{as if} it had the \textit{form} of a KLM BS with (now complex) \tit{fictitious reflection} coefficients $r_i$. 
The simplest solution was found \cite{Alsing_Hach:2019,Alsing_Hach:2018} (mimicking a calculation by Skaar \cite{Skaar:2004})  by considering the case when 
all the $\theta_i=0$ (i.e. all MRRs on resonance) and all the bus phase delays were also zero $\delta_i=0$, 
so that all the \textit{KLM effective reflection coefficients} $r_i$ were now real. 
The $S$ matrix for the MRR NLPSG taking $r_3=r_1$ is given by \cite{Alsing_Hach:2019,Alsing_Hach:2018}
\be{SMRR}
\hspace{-.2in}
S^{(MRR)} = 
\dfrac{1}{1-\,(1-r_1^2)\,r_2}\,
\left(\begin{array}{ccc}
(1-r_1^2)-r_2                                    & r_1\,\sqrt{1-r_2^2}        &   -r_1\,\sqrt{1-r_1^2}\, \sqrt{1-r_2^1} \\
r_1\,\sqrt{1-r_2^2}                            &   r_1^2\,r_2                   & \sqrt{1-r_1^2}\,(1-r_2)                       \\
 -r_1\,\sqrt{1-r_1^2}\, \sqrt{1-r_2^1} &   \sqrt{1-r_1^2}\,(1-r_2) & r_1^2
\end{array}\right).
\ee
The form of $S^{(MRR)}$ now differs from that of the  KLM case, only because in the MRR case, the middle photon, mode-$2$ runs \textit{backwards} (right to left), and so there is some involved mode-swap algebra \cite{Alsing_Hach:2019,Alsing_Hach:2018} that takes place in forming $S^{(MRR)}$  from $S^{(KLM)}$. As such, it is more compact to write $S^{(MRR)}$ as a function of both $r_1$ and $r_2$.
Analogous to \Eq{r1:of:r2} and \Eq{beta:of:r2}  we find by imposing only Condition-0 and Condition-2 we have
\bsub
\bea{t1:eq:t3:of:t2:MRR}
r_1(r_2) = r_3(r_2) &= &
\left[
\dfrac{(1-r_2)}{r_2\,(1+r_2^2)} \, \left( (1+2\,r_2-r_2^2) \mp (1+r_2)\,\sqrt{(1-3\,r^2_2)} \right)
\right]^{1/2}, 
\eea
giving rise to 
\bea{beta:of:t2:MRR}
\beta(r_2) &=&
\left( 
\dfrac{1}{2\pm \sqrt{1-3\,r_2^2} }
\right)\,
\left[
\dfrac{1+2\,r_2-r_2^2}{1+r_2} \pm \sqrt{1-3\,r_2^2} 
\right],        
\eea
\esub
where in both \Eq{t1:eq:t3:of:t2:MRR} and  \Eq{beta:of:t2:MRR} 
the top sign corresponds to the region $-1/\sqrt{3}\le r_2 \le -1/2$, and 
the bottom sign to the region $0\le r_2 \le 1/\sqrt{3}$,
which are inequivalent solutions.
By additionally imposing Condition-1 and maximizing over $r_2$ we find analogous to \Eq{r1:r2:beta:star:KLM}
\bsub
\bea{r1:r2:beta:star:MRR}
                               r_2^*     &=& \dfrac{1+2\,\sqrt{2}}{7} \hspace{.35in}= 0.546918,         \qquad r_2^{* 2} = 0.299119, \\
\trm{MRR:}\qquad  r^*_1=r^*_3 &=& \sqrt{2\,(\sqrt{2}-1)}\;\; = 0.91018, \qquad \;r_1^{* 2} =  r_3^{* 2} = 0.844778, \\
                          \beta  &=& \dfrac{1}{2},                                                      \qquad \hspace{1.5in}|\beta|^2 = \dfrac{1}{4}, 
\eea
\esub
with maximum NLPSG success probability $|\beta|^2_{max} = 1/4$.

\begin{figure}[ht]
\begin{tabular}{c}
\hspace{1in}
\includegraphics[width=4.0in,height=2.5in]{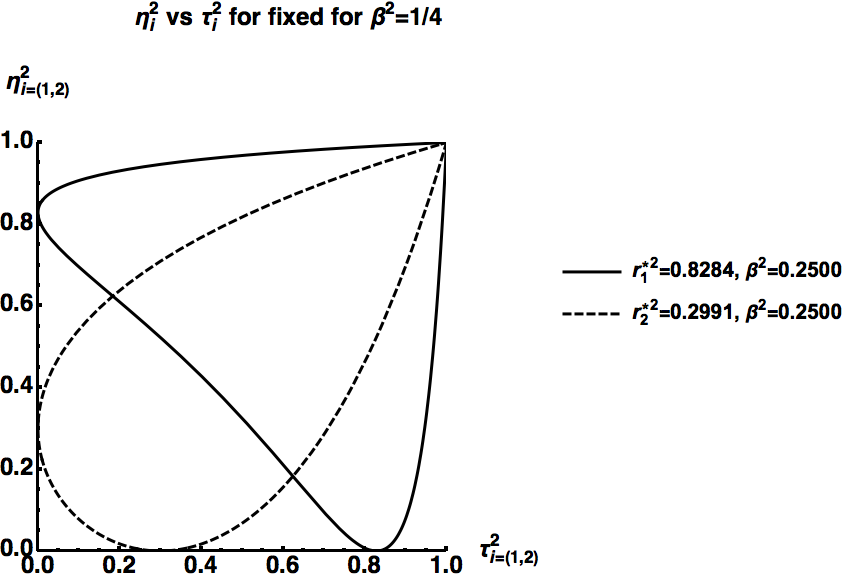}
\end{tabular}
\caption{
Plots of the physical MRR  transmissivities  
(solid) $\eta^2_{1=3}$ vs $\tau^2_{1=3}$, and
(dashed) $\eta^2_{2}$ vs $\tau^2_{2}$
for \textit{fixed} values of the 
fictitious KLM reflectivities $r^{*2}_2$ yielding $\beta^2=1/4$.  
(\Fig{fig:etaSqrd:tauSqrd:fixed:t2Sqrd:betaSqrd:in:body:of:manuscript} repeated here for clarity).
}\label{fig:etaSqrd:tauSqrd:fixed:t2Sqrd:betaSqrd:in:body:of:appendix}
\end{figure}

The difference between the analysis in \cite{Alsing_Hach:2019,Alsing_Hach:2018} and in this present work
is that  here we \textit{also} want to consider the case of  cl-SPDC input states
(only containing the states  $\ket{0}_{k\in\{1,4\}}$ and  $\ket{2}_{k\in\{1,4\}}$), 
in addition to the full w-CS input states (also containing the states $\ket{1}_{k\in\{1,4\}}$). 
In the case of cl-SPDC input states, we will find that \textit{not} imposing Condition-1, namely not letting $\beta_1$ to be equal necessarily to $\beta$ (now defined by imposing \textit{only} Condition-0 and Condition-2) 
gives qualitatively the same coincidence interference curves.
%
\textit{More importantly}, the MRR solutions for the fictitious KLM reflections coefficients $r^*_i$ 
define a 1-parameter family of physical \textit{transmission} coefficients
$\eta_i~=~\eta_i(\tau_i; r^*_i)$  regardless of the value of $ |\beta|^2$ associated with the chosen value of $r^*_i$ 
 \be{eta:tau:t:summary:app}
 \eta_i(\tau_i; r^*_i)= \dfrac{r^*_i+ \tau_i}{1+r^*_i\, \tau_i}, \quad  |\tau_i| \le 1 \Rightarrow  |\eta_i| \le 1,
  \quad \textrm{for fixed} \quad  |r^*_i| \le 1, 
 \ee 
 as illustrated in \Fig{fig:etaSqrd:tauSqrd:fixed:t2Sqrd:betaSqrd:in:body:of:appendix} 
 for values of $r_1^{*2}$ and $r_2^{*2}$ yielding $\beta^2=1/4$ for both the KLM-NLPSG and MRR-NLPSG.
 %
 The analysis for the case of MRR runs similarly for the KLM case by  merely replacing $S^{(KLM)}\to S^{(MRR)}$
 in the unitary matrix $\U$. The values of $r^*_i$ now differ from the KLM case, only because in the MRR case, the middle photon, mode-$2$ runs \textit{backwards}, and so there is some mode-swap algebra that takes place in forming $S^{(MRR)}$  from $S^{(KLM)}$. 

 \section{Action of the BS on $\ket{n,m}_{ab}$}\label{app:BS}
\begin{figure}[ht]
\begin{center}
\includegraphics[width=3.5in,height=1.55in]{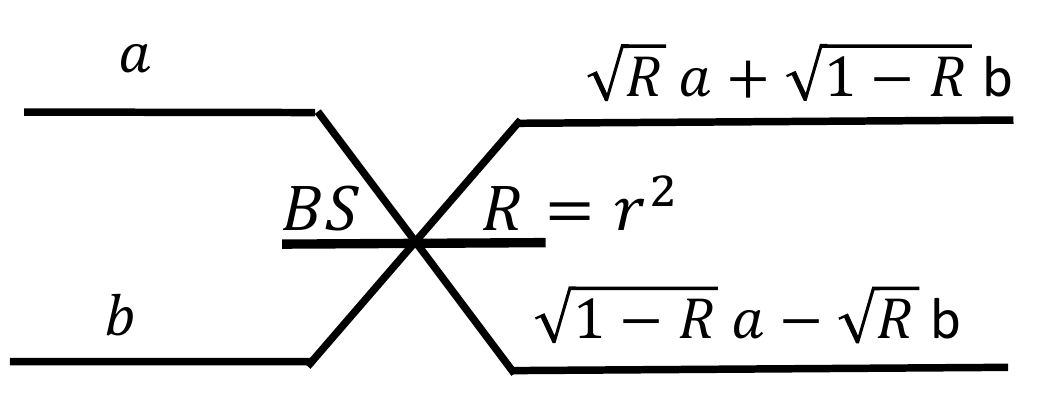}
\caption{\label{fig:BS:app} Two optical modes $a$ and $b$ mixing on a BS of reflectivity $R$. 
Note that in this (not always standard) representation the $a_{in}$ mode is the top-left input, while the $a_{out}$ mode is defined via \textit{where $a_{in}$ transmits to}, i.e. as the lower right output. Similarly for $b_{in}$ and $b_{out}$.
Here, the bottom of the BS imparts a $\pi$ phase shift of $-1$ upon reflection.
}
\end{center}
\end{figure}

 We need to know how an ideal, lossless BS acts on an arbitrary input state $\ket{n,m}$ presented at its two input ports.
 Let us define the BS transformation (Hamiltonian) on two modes $a$ and $b$ as $BS=(\theta/2)\,(a\,b^\dag + a^\dag\,b)$.
 Here $R \equiv \sin^2(\theta/2)$ is the reflectivity and  $T = (1-R) = \cos^2(\theta/2)$ is the transmissivity, such that $R+T=1$, as shown in \Fig{fig:BS:app}. (Note: we call the quantities $\sin(\theta/2)$ and $\cos(\theta/2)$ reflection and transmission \tit{coefficients}).
 The factor of $1/2$ in the argument $\theta/2$ is introduced so that $\theta=\pi/2$ represents a 50:50 BS.
%

 The action of the BS on an arbitrary input of Fock states 
$\ket{n}_a\ket{m}_b \equiv \dfrac{(a^\dag)^n}{\sqrt{n!}}\,\dfrac{(b^\dag)^n}{\sqrt{m!}}\,\ket{0}_a\ket{0}_b$ is straightforwardly computed 
(see Chapter 5 of Agarwal \textit{Quantum Optics} \cite{Agarwal:2013})
by applying the BS transformation to the last expression, and expanding out terms using the binomial theorem (since $a^\dag$ and $b^\dag$ commute).
Note that if we write the BS transformation $S_{BS}$ of the \textit{out} operators in terms of the \textit{in} operators as
$\vec{a}^\dag_{out} = S_{BS}\, \vec{a}^\dag_{in}$
then to transform an input state such as $\ket{1}_a\,\ket{0}_b = a^\dag_{in}\ket{0}_a\,\ket{0}_b$, we need to
write the \textit{in} operators in terms of the \textit{out} operators using the \textit{transpose} transformation $S^T_{BS}$ as
$\vec{a}^\dag_{in} = S^T_{BS}\, \vec{a}^\dag_{out}$ via
\be{out:in:and:in:out}
\vec{a}^\dag_{out}=
\left[
\begin{array}{c}
a^\dag_{out} \\
b^\dag_{out}
\end{array}
\right] = 
\left[
\begin{array}{cc}
\cos(\theta/2) & -\sin(\theta/2)\\
\sin(\theta/2) & \cos(\theta/2)
\end{array}
\right]
\,
\left[
\begin{array}{c}
a^\dag_{in} \\
b^\dag_{in}
\end{array}
\right] 
\equiv
S_{BS}\, \vec{a}^\dag_{in}
\Rightarrow \vec{a}^\dag_{in} = S^T_{BS}\, \vec{a}^\dag_{out}.
\ee
Thus, for example
$a^\dag_{in}\ket{0}_a\,\ket{0}_b \to \left(\cos(\theta/2)\,a^\dag_{out} + \sin(\theta/2)\,b^\dag_{out}\right)\, \ket{0}_a\,\ket{0}_b 
= \cos(\theta/2)\,\ket{1}_a\,\ket{0}_b + \sin(\theta/2)\ket{0}_a\,\ket{1}_b$. We can drop all the  \textit{in, out} labels and just remember to use the transformation $S^T_{BS}$ in computing the BS transformation formula.
The derivation is easily carried out (see also Agarwal \cite{Agarwal:2013}) with the results given below 
using $S^T_{BS}$ to transform an input state $\ket{n}_a\ket{m}_b$ to an output state, yielding
\bea{BS:Agarwal}
&{}& \ket{n}_a\ket{m}_b \to \sum_{p=0}^{n+m}\, f^{(n,m)}_p \,\ket{p}_a\,\ket{n+m-p}_b, \label{BS:on:n:m}, \\
&{}& f^{(n,m)}_p = \sum_{q=0}^n  \sum_{q'=0}^m \delta_{p,q+q'}\, \binom{n}{q}\,\binom{m}{q'}\,
\sqrt{\dfrac{p!\,(n+m-p)!}{n!\,m!}}\,
\left(-1\right)^{q'}\,
   \left(\cos(\theta/2)\right)^{m+q-q'}
\, \left(\sin(\theta/2)\right)^{n-q+q'}. \qquad  
\eea
Note that the delta function $\delta_{p,q+q'}$ ensures that the BS mixes the original input state $\ket{n}_a\ket{m}_b$ only amongst the
$n+m+1$ states of total photon number $n+m$ of the form 
$\{ \ket{0}_a\ket{n+m}_b,\, \ket{1}_a\ket{n+m-1}_b, \ldots  \ket{n}_a\ket{m}_b,\ldots,   \ket{n+m-1}_a\ket{1}_b,\  \ket{n+m}_a\ket{0}_b\}$.
The (real) BS coefficients $f^{(n,m)}_p$ are easily worked out by hand by considering states  
$\ket{n}_a\ket{m}_b$ up to $n+m=2$ at the input ports of BS, namely: 
\bsub
\bea{fp:n:0:1}
p=0: &{} \no
\ket{0}_a\ket{0}_b &\to& \ket{0}_a\ket{0}_b \Rightarrow  f^{(0,0)}_0 = 1, \\
p=1: &{} \no
\ket{0}_a\ket{1}_b &\to&  [\cos(\theta/2)\,b - \sin(\theta/2)\,a ]\ket{0}_a\ket{0}_b, \no
                   					          &=&  \cos(\theta/2) \ket{0}_a\ket{1}_b - \sin(\theta/2) \ket{1}_a\ket{0}_b, 
					                           \Rightarrow   \left\{ \begin{array}{ccr}  f^{(0,1)}_0 &=& \cos(\theta/2) \\
					                                                                                       f^{(0,1)}_1 &=& -\sin(\theta/2) 
					                                                          \end{array}  
					                                                 \right. \\ 
\ket{1}_a\ket{0}_b &\to&   [\cos(\theta/2)\,a + \sin(\theta/2)\,b ]\ket{0}_a\ket{0}_b, \no
                   					          &=&   \sin(\theta/2) \ket{0}_a\ket{1}_b + \cos(\theta/2) \ket{1}_a\ket{0}_b, 
					                           \Rightarrow   \left\{ \begin{array}{ccr}  f^{(1,0)}_0 &=& \sin(\theta/2), \\
					                                                                                      f^{(1,0)}_1 &=&   \cos(\theta/2), 
					                                                          \end{array}  
					                                                 \right. \\
\eea
\bea{fp:n:2}
p=2: &{} \no
\ket{0}_a\ket{2}_b &=&  \dfrac{1}{\sqrt{2}}\, b^2\,\ket{0}_a\ket{0}_b \to  \dfrac{1}{\sqrt{2}}\,[\cos(\theta/2)\,b - \sin(\theta/2)\,a ]^2\ket{0}_a\ket{0}_b, \no
                   					          &=&  \cos^2(\theta/2) \ket{0}_a\ket{2}_b -\sin(\theta) \ket{1}_a\ket{1}_b +  \sin^2(\theta/2) \ket{2}_a\ket{0}_b, 
					                           \Rightarrow   \left\{ \begin{array}{ccl}  f^{(0,2)}_0 &=& \cos^2(\theta/2), \\
					                                                                                       f^{(0,2)}_1 &=& -\frac{1}{\sqrt{2}}\,\sin(\theta),  \\
					                                                                                       f^{(0,2)}_2  &=& \sin^2(\theta/2), 
					                                                          \end{array}  
					                                                 \right. \\ 			                                                   
\ket{1}_a\ket{1}_b   &=& a\,b\,\ket{0}_a\ket{0}_b \to [\cos(\theta/2)\,a + \sin(\theta/2)\,b ]\,[\cos(\theta/2)\,b - \sin(\theta/2)\,a ]\ket{0}_a\ket{0}_b, \no
                   					          &=&  \sqrt{2}\,\sin(\theta/2)\,\cos(\theta/2) \ket{0}_a\ket{2}_b 
					                                    [\cos^2(\theta/2) - \sin^2(\theta/2)] \ket{1}_a\ket{1}_b , \no
					                                    & & \hspace{1.75in} - \sqrt{2}\sin(\theta/2)\,\cos(\theta/2) \ket{2}_a\ket{0}_b, 
					                           \Rightarrow   \left\{ \begin{array}{ccl}  f^{(1,1)}_0 &=& \frac{1}{\sqrt{2}} \sin(\theta), \\
					                                                                                       f^{(1,1)}_1 &=&                             \cos(\theta),  \\
					                                                                                       f^{(1,1)}_2  &=& -\frac{1}{\sqrt{2}} \sin(\theta), 
					                                                          \end{array}  
					                                                 \right. \\ 
\ket{2}_a\ket{0}_b  &=&  \dfrac{1}{\sqrt{2}}\, a^2\,\ket{0}_a\ket{0}_b \to \dfrac{1}{\sqrt{2}}\, [\cos(\theta/2)\,a + \sin(\theta/2)\,b ]^2\ket{0}_a\ket{0}_b, \no
                   					          &=&  \sin^2(\theta/2) \ket{0}_a\ket{2}_b +\sin(\theta) \ket{1}_a\ket{1}_b +  \cos^2(\theta/2) \ket{2}_a\ket{0}_b, 
					                           \Rightarrow   \left\{ \begin{array}{ccl}  f^{(2,0)}_0 &=& \sin^2(\theta/2), \\
					                                                                                       f^{(2,0)}_1 &=& \frac{1}{\sqrt{2}}\,\sin(\theta),  \\
					                                                                                       f^{(2,0)}_2  &=& \cos^2(\theta/2), 
					                                                          \end{array}  
					                                                 \right. \qquad					                                                 
\eea
\esub
Note: for each $(n,m)$ we have $\sum_{p=0}^{n+m} | f^{(n,m)}_p|^2 = 1$, which just indicates that the BS transformation is unitary.
Note  that the $f^{(n,m)}_p$  are just the Wigner rotation coefficients for the representation of a system with spin $J = (n+m)/2$ in the angular momentum basis
$\ket{J,M}$ with $2 J + 1 = n+m+1$ states $M\in\{-J, -J+1,\ldots,J\}$ where $M(p) = -J + p\, (2 J) / (n+m)$ for $p\in\{0,\ldots,n+m\}$.  

 \section{$\U$ coefficients for the 4- and 5-photon accidental states}\label{app:U:coeffs}
\subsection{The 5-photon accidental states}
 The 5-photon state  $\ket{T_{2,2'}}_{1234}$ in \Eq{Psi:2:again} proportional to $\a_2\,\a'_2\,e^{i\,2\,\varphi}$ contributes \textit{accidentals} (noise terms) to the primary coincidence counts by transferring (rerouting) photons \textit{into}  states that will be counted as coincidence counts under finite detection efficiencies. These states arise via the BS interaction on mode $1$ and $4$.
These \textit{accidentals} states \textit{do not} contribute to the primary interference terms since they are all part of $\ket{\psi^\perp_{out}}$ and hence are orthogonal to $\ket{1,1,0,1}_{1234}$ upon which the primary coincidence interference effect takes place. Further, since each orthogonal state is multiplied by an \textit{overall} phase factor $e^{i\,2\,\varphi}$, this phase factor squares to unity in the final probability sum, and hence does not even interfere, in higher order, with other states in $\ket{\psi^\perp_{out}}$. Note also that these accidentals involve states with a total photon number of $5$, while the primary coincidence interfering terms contain a total of $3$ photons. So our approximation would see a pure coincidence interference pattern if we were to stop at the $3$-photon level. However, since our initial input state is already a $5$-photon state, a reasonable \textit{self consistent, lowest order} calculation would be to consider states with up to $5$ photons, as we do here.

The total number of $5$-photon Fock states in $\ket{T_{2,2'}}$ \Eq{Psi:2:again} is $4^3\times 3=192$ (4 creation operators in each of 3 sums, and 3 terms from expanding the square of the BS operation on mode 4).
To get a handle on what terms to keep, it is useful to indicate the possible boson creation operator indices $a^\dag_j\,a^\dag_k\,a^\dag_\ell$
as $(j) [k,\ell] (1,1)$,  $(j) [k,\ell] (4,4)$ and $(j) [k,\ell] (1,4)$ where the last set of indices in parentheses indicate the terms $(a^\dag_1)^2$, $(a^\dag_4)^2$ and $a^\dag_1\,a^\dag_4$ from the expansion of the BS on mode-$4$. Consider the first set of indices $(1) [k,\ell] (11)$. These can be completely eliminated since it contains three $1$s corresponding to a state $\ket{3,\cdot,\cdot,\cdot}_{1234}$ which is outside our approximation which keeps terms with at most two photons in any single mode.

For the next set of indices $(2) [k,\ell] (11)$ we observe that since there already exists two $1$s and one $2$, the indices $[k,\ell]$ cannot contain a $1$ (since that would give three photons in mode-$1$), nor can it contain $[k,\ell] = [2,2]$ (since that would give a state with three photons in mode-$2$). 
Further, the contributing indices \textit{must} contain as a subset, the indices $\{1,2,4\}$ since terms that don't are multiplied by $p^{(1,2,4)}_0=0$.
Thus the $5$ contributing index sets are given by 
$(2) \{ [2,4], [3,4], [4,2], [4,3], [4,4] \} (11)$ corresponding to state 
$\{
\ket{2,2,0,1}_{1234}, 
\ket{2,1,1,1}_{1234},
\ket{2,2,0,1}_{1234},
\ket{2,1,1,1}_{1234},
\ket{2,1,0,2}_{1234},
\}$, respectively 
(e.g. $(2) [2,4] (11)$ is read off as (11) two photons in mode-$1$, (22) two photons in mode-$2$, and (4) one photon in mode-$4$).

Note that the next set of indices in line $(3) [k,\ell] (11)$ only contains two terms $[k,\ell] = \{ [2,4], [4,2]\}$ 
since all dropped terms either do not contain $\{1,2,4\}$, or contains $[k,\ell]=[3,3]$ which yields three photons in mode-$3$.

Similar to the prior case, the set of contributing indices for $(4) [k,\ell] (11)$ are
$(4) \{ [2,2], [2,3], [2,4], [3,2], [4,2] \} (11)$ corresponding to state 
$\{
\ket{2,2,0,1}_{1234}, 
\ket{2,1,1,1}_{1234},
\ket{2,1,0,2}_{1234},
\ket{2,1,1,1}_{1234},
\ket{2,1,0,2}_{1234},
\}$, respectively.
We can proceed similarly with the $(j) [k,\ell] (44)$ and  $(j) [k,\ell] (14)$, noting right off the bat that 
we can eliminate the set of indices $(4) [k,\ell] (44)$ since it contains three photons in mode-$4$.
The process is an exercise in tedious bookkeeping, but the procedure is straightforward, and yields
(note: $\{cc, cs, ss\}$ superscripts indicate terms are multiplied by $\{\cos^2(\theta/2),2\,\cos(\theta/2)\sin(\theta/2),\sin^2(\theta/2)\}$)
\bea{T:2:2prime}
\ket{T_{2,2'}}_{1234} &=&
\a_2\,\a'_2\,e^{i\,2\,\varphi} \times \no
&{}&
\ket{1,2,0,2}_{1234}\,
\left[
 \sin(\theta)\, \Ucs_{202}  +  \sin^2(\theta/2)\,\Uss_{202}
\right], \no
&+&
\ket{1,2,1,1}_{1234}\,
\left[
 \sin(\theta)\, \Ucs_{211} 
\right], \no
&+&
\ket{1,1,1,2}_{1234}\,
\left[
 \sin(\theta)\, \Ucs_{112}  +  \sin^2(\theta/2)\,\Uss_{112}
\right], \no
&+&
\ket{1,1,2,1}_{1234}\,
\left[
 \sin(\theta)\, \Ucs_{121} 
\right], \no
&+&
\ket{2,2,0,1}_{1234}\,
\left[
 \cos^2(\theta/2)\,\Ucc_{201} + \sin(\theta)\, \Ucs_{201} 
\right], \no
&+&
\ket{2,1,1,1}_{1234}\,
\left[
 \cos^2(\theta/2)\,\Ucc_{111} + \sin(\theta)\, \Ucs_{111} 
\right], \no
&+&
\ket{2,1,0,2}_{1234}\,
\left[
 \cos^2(\theta/2)\,\Ucc_{102} + \sin(\theta)\, \Ucs_{102}  +  \sin^2(\theta/2)\,\Uss_{102}
\right], 
\eea
where we have defined the coefficients
of the $\ket{1}_1 \otimes \ket{\cdot, \cdot, \cdot}_{234}$ terms as
\bsub
\bea{U:mode-1:1-photon:terms}
\hspace{-.6in}
\Ucs_{202} &=& \left( \U_{21}\,\U_{21}\,\U_{42} +  \U_{21}\,\U_{41}\,\U_{22}  \right), \\
\Ucs_{211} &=&  \frac{1}{\sqrt{2}}\,\left( \U_{21}\,\U_{31}\,\U_{22} +  \U_{31}\,\U_{21}\,\U_{22} \right), \\
\Ucs_{112} &=& \frac{1}{\sqrt{2}}\,\left( \U_{21}\,\U_{31}\,\U_{42} +  \U_{21}\,\U_{41}\,\U_{32} + 
                                                              \U_{31}\,\U_{21}\,\U_{42} +  \U_{31}\,\U_{41}\,\U_{22} + 
                                                              \U_{41}\,\U_{21}\,\U_{32} +  \U_{41}\,\U_{31}\,\U_{22} \right), \qquad\; \\
\Ucs_{121} &=& \frac{1}{\sqrt{2}}\,\left(\U_{21}\,\U_{31}\,\U_{32} +  \U_{31}\,\U_{21}\,\U_{32} + \U_{31}\,\U_{31}\,\U_{22}\right), \\
\Uss_{112} &=&  \left( \U_{11}\,\U_{21}\,\U_{32} +  \U_{11}\,\U_{31}\,\U_{22} +  \U_{21}\,\U_{11}\,\U_{32} +  \U_{21}\,\U_{31}\,\U_{12} \right), \\
\Uss_{202} &=&  \left( \U_{31}\,\U_{21}\,\U_{22} +  \U_{21}\,\U_{11}\,\U_{22} +  \U_{21}\,\U_{21}\,\U_{12}  \right),
\eea
\esub
with $\U$ defined from \Eq{U}.
(Note that the second indices of the triple products of $\U$s are always in the order $\{1,1,2\}$;
the first set of indices $\{j,k,\ell\}$ are associated with the state $\ket{j,k,\ell}_{234}$)
Similarly, the coefficients
of the $\ket{2}_1\otimes \ket{\cdot, \cdot, \cdot}_{234}$ terms are given by
\bsub
\bea{U:mode-1:2-photon:terms}
\hspace{-.6in}
\Ucc_{201} &=& \left( \U_{21}\,\U_{21}\,\U_{24} +  \U_{21}\,\U_{41}\,\U_{22} +  \U_{41}\,\U_{21}\,\U_{22}  \right), \\
\Ucc_{111} &=& \frac{1}{\sqrt{2}}\,\left( \U_{21}\,\U_{31}\,\U_{42} +  \U_{21}\,\U_{41}\,\U_{32} +  \U_{41}\,\U_{31}\,\U_{22}   \right), \\
\Ucc_{102} &=& \left( \U_{21}\,\U_{41}\,\U_{42} +  \U_{41}\,\U_{41}\,\U_{22} +  \U_{41}\,\U_{41}\,\U_{22}   \right), \\
\Ucs_{201} &=& \left( \U_{11}\,\U_{21}\,\U_{22} +  \U_{21}\,\U_{11}\,\U_{22} +  \U_{21}\,\U_{21}\,\U_{12}   \right), \\
\Ucs_{111} &=&  \frac{1}{\sqrt{2}}\,\left( \U_{11}\,\U_{21}\,\U_{32} +  \U_{11}\,\U_{31}\,\U_{22}\right) 
                                                  +  \left( \U_{21}\,\U_{11}\,\U_{32}  +  \U_{11}\,\U_{21}\,\U_{32} +  \U_{21}\,\U_{31}\,\U_{12} \right), \no
                       &+&  \frac{1}{\sqrt{2}}\,\left(   \U_{31}\,\U_{11}\,\U_{22}  +  \U_{31}\,\U_{21}\,\U_{12}                    \right), \\
\Ucs_{102} &=& \left( \U_{11}\,\U_{21}\,\U_{42} +  \U_{11}\,\U_{41}\,\U_{22} +  \U_{21}\,\U_{11}\,\U_{42} +  \U_{21}\,\U_{41}\,\U_{12}
                     +            \U_{41}\,\U_{11}\,\U_{22} +  \U_{41}\,\U_{21}\,\U_{12} \right), \qquad \\
\Uss_{102} &=& \left( \U_{11}\,\U_{11}\,\U_{22} +  \U_{11}\,\U_{21}\,\U_{12} +  \U_{21}\,\U_{11}\,\U_{12}   \right).
\eea
\esub

\subsection{The 4-photon accidental states when $\ket{1}_1$ and $\ket{1}_4$ are included in the input states}
Following the same procedure as in the previous section, the  coincidence state after projection,
 will be
(note: $\{cc, cs, ss,\}$ superscripts indicate terms are multiplied by $\{\cos^2(\theta/2),2\,\cos(\theta/2)\sin(\theta/2),\sin^2(\theta/2)\}$
and $\{c, s\}$ superscripts indicate terms are multiplied by $\{\cos(\theta/2),2\,\sin(\theta/2)\}$)
\bea{T:1:1prime:1:2prime:2:1prime}
 \ket{T_{1,1'}}_{1234} &+& \ket{T_{1,2'}}_{1234}  + \ket{T_{2,1'}}_{1234} = \no
&{}&
\hspace{-.8in}
\a_1\,\a'_1\,e^{i\,\varphi}\,
\ket{1,1,0,1}_{1234}\,
\left[
\cos(\theta/2)\,\Ucprime_{101} +  \sin(\theta/2)\,\Usprime_{101}\,
\right]
\no
&{}&
\hspace{-.8in}
+\,
e^{i\,\varphi}\,\ket{1,2,0,1}_{1234}\,
\left[
\a_1\,\a'_2\,e^{i\,\varphi}\,
\sin(\theta)\, \Ucsprime_{201}  
+
\a_2\,\a'_1\,
\left( \cos(\theta/2)\, \Ucprime_{201}  +  \sin(\theta/2)\,\Usprime_{201}\right)
\right], 
\no
&{}&
\hspace{-.8in}
+\,
e^{i\,\varphi}\,\ket{1,1,1,1}_{1234}\,
\left[
\a_1\,\a'_2\,e^{i\,\varphi}\,
\sin(\theta)\, \Ucsprime_{111}  
+
\a_2\,\a'_1\,
\left( \cos(\theta/2)\, \Ucprime_{111}  +  \sin(\theta/2)\,\Usprime_{111}\right)
\right], 
\no
&{}&
\hspace{-.8in}
+\,
e^{i\,\varphi}\,\ket{1,1,0,2}_{1234}\,
\left[
\a_1\,\a'_2\,e^{i\,\varphi}\,
\left( \sin(\theta)\, \Ucsprime_{102}  +  \cos^2(\theta/2)\,\Uccprime_{102}\right)
+
\a_2\,\a'_1\,
\left( \cos(\theta/2)\, \Ucprime_{102}  +  \sin(\theta/2)\,\Usprime_{102}\right)
\right], 
\no
&{}&
\hspace{-.8in}
+\,
e^{i\,\varphi}\,\ket{2,1,0,1}_{1234}\,
\left[
\a_1\,\a'_2\,e^{i\,\varphi}\,
\left( \sin(\theta)\, \Ucsprime_{101}  +  \sin^2(\theta/2)\,\Ussprime_{101}\right)
+
\a_2\,\a'_1\,
\left( \cos(\theta/2)\, \Ucprime_{101}  +  \sin(\theta/2)\,\Usprime_{101}\right)
\right], \qquad\;\;
\eea
where
(Note: 
all double products $\U\,\U$   have the second indices in the order $\{1,2\}$, while again
all triple products $\U\,\U\,\U$ have the second indices in the order $\{1,1,2\}$)
\bsub
\bea{Uprime:coeffs}
\Ucprime_{101} &=& \U_{11}\,\U_{22} + \U_{21}\,\U_{12} = \cos(\theta/2)\,(S_{11}\,S_{22} + S_{21}\,S_{12}) \equiv  \cos(\theta/2)\,\beta_1, \\
 \Usprime_{101} &=& \U_{21}\,\U_{42} + \U_{41}\,\U_{22} = -\sin(\theta/2)\,(S_{21}\,S_{22} + S_{11}\,S_{22})\equiv  -\sin(\theta/2)\,\beta_1, \\
\Ucsprime_{201}  &=&   \U_{21}\,\U_{22}  \\
\Ucsprime_{111}  &=&  \dfrac{1}{\sqrt{2}}\,\left(  \U_{21}\,\U_{32} +  \U_{31}\,\U_{22} \right), \\
\Ucsprime_{102}  &=& \U_{21}\,\U_{42} + \U_{41}\,\U_{22} = -\sin(\theta/2)\,(S_{21}\,S_{12} + S_{11}\,S_{22})\equiv   -\sin(\theta/2)\,\beta_1, \\
\Ucsprime_{101}  &=&  \U_{11}\,\U_{22} + \U_{21}\,\U_{12} = \cos(\theta/2)\,(S_{11}\,S_{22} + S_{21}\,S_{12}) \equiv  \cos(\theta/2)\,\beta_1, \\
\Uccprime_{102} &=&  \U_{11}\,\U_{22} + \U_{21}\,\U_{12} = \cos(\theta/2)\,(S_{11}\,S_{22} + S_{21}\,S_{12}) \equiv  \cos(\theta/2)\,\beta_1, \\
\Ussprime_{101} &=&\U_{21}\,\U_{42} + \U_{41}\,\U_{22} = -\sin(\theta/2)\,(S_{21}\,S_{12} + S_{11}\,S_{22}) \equiv  -\sin(\theta/2)\,\beta_1, \\
\Ucprime_{201}  &=&  \U_{11}\,\U_{21}\,\U_{22} + \U_{21}\,\U_{11}\,\U_{22} + \U_{21}\,\U_{21}\,\U_{12} , \\
\Ucprime_{111}  &=&  \dfrac{1}{\sqrt{2}}\,\left(  \U_{11}\,\U_{21}\,\U_{32} + \U_{11}\,\U_{31}\,\U_{22} + \U_{21}\,\U_{11}\,\U_{32}  +
                                                                          \U_{21}\,\U_{31}\,\U_{12} + \U_{31}\,\U_{11}\,\U_{22} + \U_{31}\,\U_{21}\,\U_{12} \right),\qquad\; \\
\Ucprime_{102}  &=&  \U_{11}\,\U_{21}\,\U_{42} + \U_{11}\,\U_{41}\,\U_{22} + \U_{21}\,\U_{11}\,\U_{42} + 
                                   \U_{21}\,\U_{41}\,\U_{12} + \U_{41}\,\U_{11}\,\U_{22} + \U_{41}\,\U_{21}\,\U_{12} \\
\Ucprime_{101}  &=&  \U_{11}\,\U_{11}\,\U_{22} + \U_{11}\,\U_{21}\,\U_{12} + \U_{21}\,\U_{11}\,\U_{12} , \\
\Usprime_{201}  &=&  \U_{21}\,\U_{21}\,\U_{42} + \U_{21}\,\U_{41}\,\U_{22} + \U_{41}\,\U_{21}\,\U_{22} , \\
\Usprime_{111}  &=&  \dfrac{1}{\sqrt{2}}\,\left(  \U_{21}\,\U_{31}\,\U_{42} + \U_{21}\,\U_{41}\,\U_{32} + \U_{31}\,\U_{21}\,\U_{42}  +
                                                                          \U_{31}\,\U_{41}\,\U_{22} + \U_{41}\,\U_{21}\,\U_{32} + \U_{41}\,\U_{31}\,\U_{22} \right),\qquad\; \\
\Usprime_{102}  &=&  \U_{21}\,\U_{41}\,\U_{42} + \U_{41}\,\U_{21}\,\U_{42} + \U_{41}\,\U_{41}\,\U_{22}, \\
\Usprime_{101}  &=&  \U_{11}\,\U_{21}\,\U_{42} + \U_{11}\,\U_{41}\,\U_{22} + \U_{21}\,\U_{11}\,\U_{42} + 
                                   \U_{21}\,\U_{41}\,\U_{12} + \U_{41}\,\U_{11}\,\U_{22} + \U_{41}\,\U_{21}\,\U_{12}.
\eea
\esub
In the above we have defined as in \Eq{beta:1}
\be{beta1}
\beta_1  \equiv S_{11}\,S_{22} + S_{21}\,S_{12} \myover{\longrightarrow}{\beta\to\beta_{max}=1/2} \, \beta_{max}=1/2, 
\quad (\beta_{max}^2=1/4),
\ee
where
$\beta_1\to \beta=1/2$ 
\textit{if} we were to additionally impose Condition-$1$ \Eq{beta:1}, which would then make $\beta^2\to\beta^2_{max}=1/4$ 
when all three Conditions-0,1,2 (Condition-$0$   \Eq{beta:0}, and Condition-$2$  \Eq{beta:2}) are imposed .

\begin{acknowledgments}
PMA, AMS, and MLF would like to acknowledge support of this work from
the Air Force Office of Scientific Research (AFOSR).
PLK and EEH would like to acknowledge support for this work was provided by 
the Air Force Research Laboratory (AFRL) Summer Faculty Fellowship Program (SFFP).
The authors wish thank Paul Kwiat  for useful discussions and helpful suggestions.
Any opinions, findings and conclusions or recommendations
expressed in this material are those of the author(s) and do not
necessarily reflect the views of Air Force Research Laboratory.
\end{acknowledgments}

\bibliography{rr_losses_refs}

\begin{thebibliography}{13}%
\makeatletter
\providecommand \@ifxundefined [1]{%
 \@ifx{#1\undefined}
}%
\providecommand \@ifnum [1]{%
 \ifnum #1\expandafter \@firstoftwo
 \else \expandafter \@secondoftwo
 \fi
}%
\providecommand \@ifx [1]{%
 \ifx #1\expandafter \@firstoftwo
 \else \expandafter \@secondoftwo
 \fi
}%
\providecommand \natexlab [1]{#1}%
\providecommand \enquote  [1]{``#1''}%
\providecommand \bibnamefont  [1]{#1}%
\providecommand \bibfnamefont [1]{#1}%
\providecommand \citenamefont [1]{#1}%
\providecommand \href@noop [0]{\@secondoftwo}%
\providecommand \href [0]{\begingroup \@sanitize@url \@href}%
\providecommand \@href[1]{\@@startlink{#1}\@@href}%
\providecommand \@@href[1]{\endgroup#1\@@endlink}%
\providecommand \@sanitize@url [0]{\catcode `\\12\catcode `\$12\catcode
  `\&12\catcode `\#12\catcode `\^12\catcode `\_12\catcode `\%12\relax}%
\providecommand \@@startlink[1]{}%
\providecommand \@@endlink[0]{}%
\providecommand \url  [0]{\begingroup\@sanitize@url \@url }%
\providecommand \@url [1]{\endgroup\@href {#1}{\urlprefix }}%
\providecommand \urlprefix  [0]{URL }%
\providecommand \Eprint [0]{\href }%
\providecommand \doibase [0]{http://dx.doi.org/}%
\providecommand \selectlanguage [0]{\@gobble}%
\providecommand \bibinfo  [0]{\@secondoftwo}%
\providecommand \bibfield  [0]{\@secondoftwo}%
\providecommand \translation [1]{[#1]}%
\providecommand \BibitemOpen [0]{}%
\providecommand \bibitemStop [0]{}%
\providecommand \bibitemNoStop [0]{.\EOS\space}%
\providecommand \EOS [0]{\spacefactor3000\relax}%
\providecommand \BibitemShut  [1]{\csname bibitem#1\endcsname}%
\let\auto@bib@innerbib\@empty
\bibitem [{\citenamefont {Knill}\ \emph {et~al.}(2001)\citenamefont {Knill},
  \citenamefont {Laflamme},\ and\ \citenamefont {Milburn}}]{KLM:2001}%
  \BibitemOpen
  \bibfield  {author} {\bibinfo {author} {\bibfnamefont {E.}~\bibnamefont
  {Knill}}, \bibinfo {author} {\bibfnamefont {R.}~\bibnamefont {Laflamme}}, \
  and\ \bibinfo {author} {\bibfnamefont {G.~J.}\ \bibnamefont {Milburn}},\
  }\href@noop {} {\bibfield  {journal} {\bibinfo  {journal} {Nature}\ }\textbf
  {\bibinfo {volume} {409}},\ \bibinfo {pages} {46} (\bibinfo {year}
  {2001})}\BibitemShut {NoStop}%
\bibitem [{\citenamefont {Knill}(2002)}]{Knill:2002}%
  \BibitemOpen
  \bibfield  {author} {\bibinfo {author} {\bibfnamefont {E.}~\bibnamefont
  {Knill}},\ }\href@noop {} {\bibfield  {journal} {\bibinfo  {journal} {Phys.
  Rev. A}\ }\textbf {\bibinfo {volume} {66}},\ \bibinfo {pages} {052306}
  (\bibinfo {year} {2002})}\BibitemShut {NoStop}%
\bibitem [{\citenamefont {R.~Okamoto}\ and\ \citenamefont
  {Takeuchi}(2011)}]{Obrien:2011}%
  \BibitemOpen
  \bibfield  {author} {\bibinfo {author} {\bibfnamefont {H.~H.}\ \bibnamefont
  {R.~Okamoto}, \bibfnamefont {J.L.O'brien}}\ and\ \bibinfo {author}
  {\bibfnamefont {S.}~\bibnamefont {Takeuchi}},\ }\href@noop {} {\bibfield
  {journal} {\bibinfo  {journal} {PNAS}\ }\textbf {\bibinfo {volume} {108}},\
  \bibinfo {pages} {10067} (\bibinfo {year} {2011})}\BibitemShut {NoStop}%
\bibitem [{\citenamefont {Scott}\ \emph {et~al.}(2019)\citenamefont {Scott},
  \citenamefont {Alsing}, \citenamefont {Smith}, \citenamefont {Fanto},
  \citenamefont {Tison},\ and\ \citenamefont {Hach~III}}]{Alsing_Hach:2019}%
  \BibitemOpen
  \bibfield  {author} {\bibinfo {author} {\bibfnamefont {R.}~\bibnamefont
  {Scott}}, \bibinfo {author} {\bibfnamefont {P.~M.}\ \bibnamefont {Alsing}},
  \bibinfo {author} {\bibfnamefont {A.}~\bibnamefont {Smith}}, \bibinfo
  {author} {\bibfnamefont {M.}~\bibnamefont {Fanto}}, \bibinfo {author}
  {\bibfnamefont {C.}~\bibnamefont {Tison}}, \ and\ \bibinfo {author}
  {\bibfnamefont {E.~E.}\ \bibnamefont {Hach~III}},\ }\href@noop {} {\bibfield
  {journal} {\bibinfo  {journal} {Phys. Rev. A}\ }\textbf {\bibinfo {volume}
  {100}},\ \bibinfo {pages} {022322} (\bibinfo {year} {2019})}\BibitemShut
  {NoStop}%
\bibitem [{\citenamefont {Alsing}\ and\ \citenamefont
  {Hach~III}(2018)}]{Alsing_Hach:2018}%
  \BibitemOpen
  \bibfield  {author} {\bibinfo {author} {\bibfnamefont {P.~M.}\ \bibnamefont
  {Alsing}}\ and\ \bibinfo {author} {\bibfnamefont {E.~E.}\ \bibnamefont
  {Hach~III}},\ }\href@noop {} {\bibfield  {journal} {\bibinfo  {journal}
  {Quant. Info Sci. and Tech. IV}\ }\textbf {\bibinfo {volume} {10803}},\
  \bibinfo {pages} {10803M} (\bibinfo {year} {2018})}\BibitemShut {NoStop}%
\bibitem [{\citenamefont {C.K.~Hong}\ and\ \citenamefont
  {Mandel}(1987)}]{HOM:1987}%
  \BibitemOpen
  \bibfield  {author} {\bibinfo {author} {\bibfnamefont {Z.~O.}\ \bibnamefont
  {C.K.~Hong}}\ and\ \bibinfo {author} {\bibfnamefont {L.}~\bibnamefont
  {Mandel}},\ }\href@noop {} {\bibfield  {journal} {\bibinfo  {journal} {Phys.
  Rev. Lett.}\ }\textbf {\bibinfo {volume} {59}},\ \bibinfo {pages} {2044}
  (\bibinfo {year} {1987})}\BibitemShut {NoStop}%
\bibitem [{\citenamefont {P.~Kok}\ and\ \citenamefont
  {Dowling}(2002)}]{Kok:2002}%
  \BibitemOpen
  \bibfield  {author} {\bibinfo {author} {\bibfnamefont {H.~L.}\ \bibnamefont
  {P.~Kok}}\ and\ \bibinfo {author} {\bibfnamefont {J.}~\bibnamefont
  {Dowling}},\ }\href@noop {} {\bibfield  {journal} {\bibinfo  {journal} {Phys.
  Rev. A}\ }\textbf {\bibinfo {volume} {65}},\ \bibinfo {pages} {5} (\bibinfo
  {year} {2002})}\BibitemShut {NoStop}%
\bibitem [{\citenamefont {Gerry}\ and\ \citenamefont
  {Knight}(2004)}]{Gerry_Knight:2004}%
  \BibitemOpen
  \bibfield  {author} {\bibinfo {author} {\bibfnamefont {C.}~\bibnamefont
  {Gerry}}\ and\ \bibinfo {author} {\bibfnamefont {P.~L.}\ \bibnamefont
  {Knight}},\ }\href@noop {} {\emph {\bibinfo {title} {Introductory Quantum
  Optics}}}\ (\bibinfo  {publisher} {Cambridge Univeristy Press,Cambridge},\
  \bibinfo {year} {2004})\BibitemShut {NoStop}%
\bibitem [{\citenamefont {Hach~III}\ \emph {et~al.}(2014)\citenamefont
  {Hach~III}, \citenamefont {Preble}, \citenamefont {Elshaari}, \citenamefont
  {Alsing},\ and\ \citenamefont {Fanto}}]{Hach:2014}%
  \BibitemOpen
  \bibfield  {author} {\bibinfo {author} {\bibfnamefont {E.~E.}\ \bibnamefont
  {Hach~III}}, \bibinfo {author} {\bibfnamefont {S.~F.}\ \bibnamefont
  {Preble}}, \bibinfo {author} {\bibfnamefont {A.~W.}\ \bibnamefont
  {Elshaari}}, \bibinfo {author} {\bibfnamefont {P.~M.}\ \bibnamefont
  {Alsing}}, \ and\ \bibinfo {author} {\bibfnamefont {M.~L.}\ \bibnamefont
  {Fanto}},\ }\href@noop {} {\bibfield  {journal} {\bibinfo  {journal} {Phys.
  Rev. A}\ }\textbf {\bibinfo {volume} {89}},\ \bibinfo {pages} {043805}
  (\bibinfo {year} {2014})}\BibitemShut {NoStop}%
\bibitem [{\citenamefont {Skaar}\ \emph {et~al.}(2004)\citenamefont {Skaar},
  \citenamefont {Escartin},\ and\ \citenamefont {Landro}}]{Skaar:2004}%
  \BibitemOpen
  \bibfield  {author} {\bibinfo {author} {\bibfnamefont {J.}~\bibnamefont
  {Skaar}}, \bibinfo {author} {\bibfnamefont {J.}~\bibnamefont {Escartin}}, \
  and\ \bibinfo {author} {\bibfnamefont {H.}~\bibnamefont {Landro}},\
  }\href@noop {} {\bibfield  {journal} {\bibinfo  {journal} {Am. J. Phys.}\
  }\textbf {\bibinfo {volume} {72}},\ \bibinfo {pages} {1385} (\bibinfo {year}
  {2004})}\BibitemShut {NoStop}%
\bibitem [{Note1()}]{Note1}%
  \BibitemOpen
  \bibinfo {note} {We graciously acknowledge Paul Kwiat for this insightful
  suggestion conveyed to us at the 1st Photons for Quantum (PfQ) Conference,
  Rochester Institute of Technology, Rochester, NY, 23-25Jan2019}\BibitemShut
  {NoStop}%
\bibitem [{Note2()}]{Note2}%
  \BibitemOpen
  \bibinfo {note} {In \cite {Alsing_Hach:2019, Alsing_Hach:2018} we labeled the
  $r_i$ in this work as $t_i$, and called the later \protect \textit
  {transmission coefficients}. We followed the calculation of Skaar \cite
  {Skaar:2004} so that the $t_i$ were in fact actually \protect \textit
  {reflection} coefficients. All the calculations and conclusions in \cite
  {Alsing_Hach:2019,Alsing_Hach:2018} are uneffected, since the KLM $t_i$
  functioned merely as parameters that defined the 1-dimensional manifold
  relationship between the \protect \textit {physical transmission}
  coefficients $\eta _i$ and $\tau _i$ of MRR$_i$, see Eq.(\ref {eta:tau:t})
  and Eq.(\ref {eta:tau:t:summary:app}).}\BibitemShut {Stop}%
\bibitem [{\citenamefont {Agarwal}(2013)}]{Agarwal:2013}%
  \BibitemOpen
  \bibfield  {author} {\bibinfo {author} {\bibfnamefont {G.~S.}\ \bibnamefont
  {Agarwal}},\ }\href@noop {} {\emph {\bibinfo {title} {Quantum Optics}}}\
  (\bibinfo  {publisher} {Cambridge University Press},\ \bibinfo {address}
  {Cambridge},\ \bibinfo {year} {2013})\BibitemShut {NoStop}%
\end{thebibliography}%
\end{document}